\documentclass{aa} 

\usepackage{hyperref}
\usepackage{graphicx}
\usepackage{txfonts}
\usepackage{varwidth}
\usepackage{xspace}
\usepackage{xcolor}

\newcommand{\corr}[1]{\textcolor{black}{#1}}
\newcommand{\fpc}[1]{\textcolor{black}{#1}}
\newcommand{\iap}[1]{\textcolor{black}{#1}}
\newcommand{\lupm}[1]{\textcolor{black}{#1}}
\newcommand{\heb}[1]{\textcolor{black}{#1}}
\newcommand{\eng}[1]{\textcolor{black}{#1}}

\newcommand{\Fermi}{\emph{Fermi}\xspace}
\usepackage{lineno}
\def\index{s}
\def\bb{{\it b}\xspace}
\def\cc{{\it c}\xspace}
\def\dd{{\it d}\xspace}
\def\dda{{\it d$_1$}\xspace}
\def\ddb{{\it d$_2$}\xspace}
\def\latdur{T$_\mathrm{90}^\mathrm{LAT}$\xspace}
\def\tp{t_\mathrm{peak}}
\def\tv{t_\mathrm{v}}
\def\alt{\lambda}

\begin{document} 
\title{Time evolution of the spectral break in the high-energy extra component of GRB~090926A}
\author{
  M.~Yassine\inst{1}
  \and
  F.~Piron\inst{1}
  \and
  R.~Mochkovitch\inst{2}
  \and
  F.~Daigne\inst{2}
}

\institute{
  Laboratoire Univers et Particules de Montpellier, Université de Montpellier, CNRS/IN2P3, Montpellier, France\\
  \email{manal.yassine@lupm.in2p3.fr, piron@in2p3.fr}
  \and
  UPMC-CNRS, UMR7095, Institut d’Astrophysique de Paris, 75014 Paris, France\\
  \email{mochko@iap.fr, daigne@iap.fr}
}

\date{Received July 15, 2016; accepted , 2016}

\abstract
  {}
   {
     \fpc{The prompt light curve of the long GRB~090926A reveals a short pulse $\sim$$10$~s after the beginning of the
     burst emission, which has been observed by the \Fermi observatory from the keV to the GeV energy domain.
     During this bright spike, the high-energy emission from GRB~090926A underwent a sudden hardening above 10~MeV, in the
     form of an additional power-law component exhibiting a spectral attenuation at a few hundreds of MeV.}
      This high-energy break has been previously interpreted in terms of gamma-ray \corr{opacity to pair creation}, and
      used to \fpc{estimate} the bulk Lorentz factor of the outflow.
     In this article, we report on a new time-resolved analysis of the GRB~090926A broad-band spectrum during its prompt
     phase, and on its interpretation in the framework of prompt emission models.
   }
   {
     We characterize the emission from GRB~090926A at the highest energies using Pass~8 data from the \Fermi Large Area
     Telescope (LAT), which offer a greater sensitivity than any data set used in previous studies of this burst,
     \corr{particularly in the 30-100~MeV energy band.}
     Then, we combine the LAT data with the \Fermi Gamma-ray Burst Monitor (GBM) in joint spectral fits in order to
     characterize the time evolution of the broad-band spectrum from keV to GeV energies. Careful attention is paid to the
     systematic effects arising from the uncertainties on the LAT response.
     Finally, we perform a temporal analysis of the light curves and we compute the variability time scales
     from keV to GeV energies during and after the bright spike.
   }
   {
     Our analysis confirms and better constrains the spectral break which has been previously reported during the bright
     spike.
     Furthermore, it reveals that the spectral attenuation persists at later times, with an increase of the
     break \heb{characteristic} energy up to the GeV domain until the end of the prompt phase.
     We discuss these results in terms of keV-MeV synchroton radiation of electrons accelerated during the dissipation of
     the \eng{jet's internal energy,} and inverse Compton emission at higher energies.
     \fpc{We interpret the high-energy spectral break as caused} by photon \corr{opacity to pair creation}.
     Requiring that all emissions are produced above the photosphere of GRB~090926A, \fpc{we compute} the bulk Lorentz
     factor of the outflow, $\Gamma$. The latter \corr{decreases from $230$ during the spike to $100$ at the end of the
     prompt emission.}
     Assuming, instead, that the spectral break reflects the natural curvature of the inverse Compton spectrum,
     lower limits corresponding to larger values of $\Gamma$ are also derived.
     Combined with the extreme temporal variability of GRB~090926A, these Lorentz
     factors lead to emission radii \corr{$R\sim10^{14}$~cm} which are consistent with an internal origin of both the
     keV-MeV and GeV prompt emissions.
   }
   {}
   \keywords{
     gamma-ray bursts -- gamma-ray \corr{opacity to pair creation} -- jet Lorentz factor -- prompt emission -- synchrotron and inverse
     Compton radiations
   }
\maketitle

\section{Introduction}
\label{sec:intro}
\fpc{To a large extent, the physical mechanims at work in Gamma-Ray Bursts (GRBs) remain elusive more than 40 years after
  their discovery.
  The current paradigm~\citep[see, e.g.,][]{piran04} associates these powerful flashes of hard X rays and gamma rays with explosions of massive stars (the
  so-called long GRBs) or with the merging of neutron stars or black hole-neutron star binaries (short GRBs).
  These events can be detected from galaxies at cosmological distances due to their huge luminosity, which is caused by an
  ultra-relativistic outflow moving towards the observer.
  Independently of the GRB progenitor, the phenomenology distinguishes two consecutive phases of non-thermal
  emissions, with different temporal properties.
  The prompt phase of short GRBs lasts typically less than 2~s, and it can continue for several minutes in some long GRBs.
  The prompt gamma-ray emission is the most intense and often highly variable, with light curves that generally exhibit
  multiple pulses at different time scales.
  This contrasts with the smoother evolution of the afterglow phase that is observed at later times, where the maximum of
  the emission cools down to the X-ray and radio domains on a daily timescale as the overall intensity decreases.}

\fpc{The physical mechanisms that are responsible for the GRB prompt emission are still highly debated.
  In the internal shock scenario, the fast variability observed at early times is caused by shocks taking
  place within the jet, which accelerate the particles in the outflow and produce non-thermal radiations.}
\corr{\citep{rees94,kobayashi97,daigne98}}.
Magnetic reconnection has been also discussed as an alternative to internal shocks in the case of outflows which are still
highly magnetized at large distance \corr{\citep{mckinney2012,zhang14,beniamini16}}.
In these models, the prompt emission is produced above the photosphere as suggested by the non-thermal spectrum.
However, it has been shown that non-thermal emission can also be produced at the photosphere if a sub-photospheric
dissipation mechanism is at work \corr{\citep{rees05,peer05,ryde11,giannios12,beloborodov13}}.
\fpc{After the prompt phase, the afterglow is produced at larger distances and is due to the interaction of the jet with
  the ambient medium, which creates a strong external shock.}\\

Since the launch of the \Fermi observatory in June 2008, the Large Area Telescope (LAT) has detected more than 100 GRBs
above 20~MeV\footnote{\tt{http://fermi.gsfc.nasa.gov/ssc/observations/types/grbs/lat\_grbs/table.php}.}~\citep{vianello2015}.
The second instrument onboard \Fermi, the Gamma-ray Burst Monitor (GBM), has \fpc{detected 1400 GRBs in the sub-MeV range
  during the first six years~\citep{bhat2016}, and more than 2000 as of today.}
Together, the GBM and the LAT have provided a wealth of new information on the temporal and spectral properties of GRBs
over a wide energy range.
\fpc{The properties of GRBs at high energies have been investigated in detail in the first LAT
  GRB catalog~\citep{latgrbcat}.
  In general, their emission above 100~MeV starts significantly later than their keV-MeV prompt emission recorded by the GBM,
  and it continues over a much longer time scale.
  When sufficient photon statistics were available, their GeV emission was also found to be harder than the
  extrapolation of their keV-MeV emission spectrum, and generally well described by a power-law spectral component with a
  photon index $\gtrsim$$-2$.
  After the end of the keV-MeV prompt emission, this additional power-law component persists during hundreds of seconds,
  up to 19~hours in the case of GRB~130427A~\citep{130427a}.
  Specifically, \cite{latgrbcat} showed that the luminosity above 100~MeV decreases simply as $\displaystyle L(t)\propto
  t^{\alt}$, with $\alt\simeq-1$ at late times.}

\fpc{A possible interpretation of these results (delayed onset of the GeV emission, power-law temporal decay of the
  long-lived GeV emission) considers the synchrotron emission from electrons accelerated at the external shock to explain
  the entire signal detected by the
  LAT~\citep{razzaque10,kumar10,ghisellini10,090510_afterglow,110731a,lemoine13,wang13}.}
There is however a theoretical argument against this interpretation, as emphasized by~\cite{beloborodov14}: \corr{the LAT
  flux usually starts to decrease well before the end of the prompt emission in the GBM, which is too early to correspond
  to the self-similar stage of the afterglow evolution, expected on theoretical grounds at somewhat later times.}
Alternative models are based on the interaction of prompt photons with the shocked and/or unshocked ambient
medium~\citep[see, e.g.,][]{beloborodov14} or imply a contribution of internal dissipation mechanisms to the LAT flux at
early times. As discussed below, such an internal contribution seems unavoidable when variability is observed in the LAT.\\

Indeed, despite its ability to account for several observed high-energy properties of GRBs, the interpretation presented
above has proven to be insufficient to explain all of the LAT GRB observations.
\fpc{The study of GRBs~090510, 090926A and 090902B by ~\cite{latgrbcat} revealed a flattening in the
  power-law temporal decay of the luminosity above 100~MeV well after the end of the keV-MeV prompt emission.
  For instance, the decay index $\alt$ of GRB~090926A increased from $\sim$$-2.7$ to $\sim$$-0.9$ at $\sim$$40$~s
  post-trigger, while the prompt emission lasted only $\sim$$22$~s in the GBM~\citep{090926a}.
  \cite{latgrbcat} interpreted this flattening as a possible evolution from a phase where internal and external emissions
  combine at GeV energies, to a phase where the afterglow emission prevails.}
\fpc{Actually}, an internal origin of the high-energy emission has to be favored during highly variable episodes, as
observed in the prompt light curve of GRB~090926A.
\fpc{It is noteworthy that the additional power-law component in the spectrum of this burst was detected at the
  time of a short and bright pulse, observed synchronously by the GBM and the LAT at $\sim$$10$~s post-trigger.
  The attenuation of this spectral component at a few hundreds of MeV} 
has been previously interpreted in terms of gamma-ray \corr{opacity to pair creation}, and used to \fpc{estimate} the bulk
Lorentz factor of the outflow~\citep{090926a}.\\
      
In this article, we reanalyse the broad-band prompt emission spectrum of GRB~090926A using LAT Pass~8 data, which offer a
greater sensitivity than any LAT data selection used in previous studies of this burst, \corr{particularly in the
  30-100~MeV energy band.}
In Sect.~\ref{sec:analysis}, we present the \Fermi/GBM and LAT data samples, and our spectral analysis methods.
In Sect.~\ref{sec:results}, we combine the GBM and LAT data in joint spectral fits in order to characterize the time
evolution of the spectrum from keV to GeV energies. Careful attention is paid to the systematic effects arising
from the uncertainties on the LAT response.
Finally, we perform a temporal analysis of the light curves and we compute the variability time scales
from keV to GeV energies during and after the bright spike.
We discuss these results in Sect.~\ref{sec:theory} in terms of keV-MeV synchroton radiation of electrons accelerated
during the dissipation of the \eng{jet's internal energy}, and inverse Compton emission at higher energies.
\fpc{We interpret the high-energy spectral break as caused by photon \corr{opacity to pair creation}. Requiring} that all
emissions are produced above the photosphere of GRB~090926A, we \fpc{estimate} the bulk Lorentz factor of the
outflow \corr{and its time evolution}.
Our conclusions are given in Sect.~\ref{sec:concl}.

\section{Data preparation and spectral analysis methods}
\label{sec:analysis}
\begin{figure*}[!t]
  \centering
  \includegraphics[width=\linewidth]{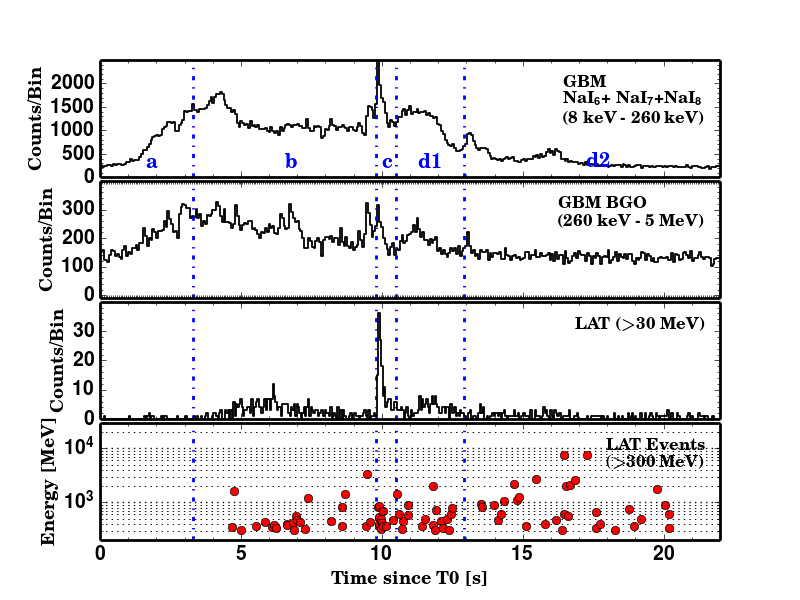}
  \caption{
    \corr{GRB~090926A counts light curves as measured by the GBM and the LAT, from lowest to highest energies: sum of the
      counts in the GBM NaI detectors (first panel), in the GBM BGO detector facing the burst (second panel), and using
      the LAT Pass~8 transient-class events above 30~MeV within a 12$^{\circ}$ region of interest (third panel). The last panel displays
      the energies of the events from this sample which have been detected above 300~MeV.
      The dashed blue vertical lines mark the time intervals which are used for the joint GBM and LAT spectral analyses.
    }
  }
  \label{fig:multidet_lc}
\end{figure*}

\subsection{Observations and data selection}
\label{subsec:data}
\fpc{GRB~090926A has been observed over a broad energy range by the two instruments onboard the \Fermi observatory,
the GBM and the LAT.
The GBM~\citep{meegan09} is a set of 12 NaI and 2 BGO scintillators installed around the spacecraft in order to cover
a large fraction of the sky.
While the onboard trigger is based on the signal recorded by the NaI detectors only, both NaI (8-1000~keV) and BGO
(0.15-40~MeV) detectors are used for spectral analyses on the ground.
The LAT~\citep{atwood09} is the main instrument of \Fermi.
This pair-conversion telescope can cover the high-energy part of GRB spectra, from 20~MeV up to more than 300~GeV.}
\fpc{GRB~090926A triggered the GBM on 2009 September 26, at T$_0=$ 04:20:26.99 UT, and it occured at an off-axis
angle of 48$^\circ$ in the LAT field of view.
The} GBM and LAT response remained essentially constant during the prompt emission phase of the burst.
Later follow-up observations of the optical afterglow of GRB~090026A placed this burst at a redshift $z=2.1062$.
\corr{Adopting a Hubble constant of
  $H_0=72$~km\,s$^{-1}$\,Mpc$^{-1}$ and cosmological parameters of $\Omega_\Lambda = 0.73$ and $\Omega_M = 0.27$ as
  in~\cite{090926a}, this corresponds to a luminosity distance $D_{\rm L}(z)=5.1\ 10^{28}$~cm.}\\


\begin{table*}[!t]
  \caption{Event statistics in Pass~7 and Pass~8 data during the \latdur of GRB~090926A (from 5.5~s to 225~s
    post-trigger).}
  \label{tab:stat}
  \centering
  \begin{tabular}{l c c c}
    \hline
    \hline
    Energy range &  Number of Pass~7 events & Number of Pass~8 events & Pass~8/Pass~7\\
    \hline  
    30~MeV-50~MeV &  33 & 243 & 7.4   \\
    50~MeV-0.1~GeV  &  95 & 381 & 4.0 \\
    0.1~GeV-0.5~GeV & 257 & 391 & 1.5 \\
    0.5~GeV-1~GeV  &  29 & 40 & 1.4 \\
    1~GeV-10~GeV &  32 & 32 & 1.0 \\
    10~GeV-100~GeV     &  1 & 1 & 1.0 \\
    Total & 447 & 1088 & 2.4 \\
    \hline
  \end{tabular}
\end{table*}

Following the analysis reported in~\cite{090926a}, we selected the GBM Time-Tagged Event (TTE) data from three NaI
detectors (N6, N7, N8) and one BGO detector (B1). The GBM TTE data are unbinned in time and have the finest time (2
$\mu$s) and energy resolution that can be reached by the GBM.
Since the launch of \Fermi, the LAT event classes have been publicly released as different versions or ``passes'' of the data,
corresponding to different instrument responses.
The results reported in~\cite{090926a} and in the first LAT GRB catalog~\citep{latgrbcat} are based on Pass~6 data.
In this work, we used the Pass~8 data which have been released in June 2015 at the \Fermi Science
Support Center~\footnote{\url{http://fermi.gsfc.nasa.gov/ssc}} (FSSC \corr{hereafter}).
These data have been processed using more elaborate reconstruction and classification algorithms.
Most importantly for the purpose of GRB analyses, the LAT effective area has been greatly improved and the spectral reach of the
instrument has been extended, with the possibility to include photons with energies lower than 100~MeV, where the gain in
effective area is the largest.
Specifically, our analysis of GRB~090926A is based on the ``P8R2$\_$TRANSIENT100$\_$V6'' event class, corresponding to
event selection cuts which have been optimized for the study of short gamma-ray transients.
In order to show how Pass~8 data improves the LAT sensitivity to GRB spectral features, we repeated part of our
analysis above 100~MeV using Pass~7 data (``P7REP$\_$TRANSIENT$\_$V15''), since the event reconstruction
between Pass~6 and Pass~7 remained essentially unchanged.
In all of our analyses, we selected the transient class events which fall in a Region of Interest (RoI) with fixed radius
of 12$^{\circ}$.
In order to avoid any residual contamination from the Earth's limb, i.e. from $\gamma$-rays produced by the interactions of
cosmic rays in the upper atmosphere, we also excluded all time intervals with a RoI zenith angle larger than $105^\circ$.\\

\corr{The GRB~090926A counts light curves based on the selections of the GBM and LAT data described above are shown in
  Fig.~\ref{fig:multidet_lc}.
  For the joint GBM and LAT spectral analyses presented further in this article, we used the three time intervals (\bb,
  \cc, \dd) which have been defined in~\cite{090926a}, with boundaries at T$_{0}$ + (3.3, 9.8, 10.5, 21.6)~s, as displayed
  in Fig.~\ref{fig:multidet_lc}.
  We ignored the data taken during the 3.3 seconds post-trigger (time interval {\it a} in Fig.~\ref{fig:multidet_lc})
  since GRB~090926A was not detected by the LAT during this period~\citep{090926a}.}
We also performed spectral analyses using LAT-only data over the whole duration of the burst.
The corresponding time interval (\latdur hereafter) that we adopted runs from T$_{0}$ + 5.5~s to T$_{0}$ + 225~s,
in accordance with the duration of the LAT emission reported in~\cite{latgrbcat} (note that this interval is much longer
than the duration of $\sim$15~s measured by the GBM).
Table~\ref{tab:stat} shows the Pass~7 and Pass~8 event statistics collected by the LAT during the \latdur time interval.
About 2.4 times more events enter the Pass~8 selection, the gain in statistics being the largest below 100~MeV, with an increase
of the event numbers by a factor of 4 to 7 depending on the energy range.

\subsection{LAT-only spectral analysis}
\label{subsec:lat_ana} 
The LAT spectral analyses were performed with the suite of standard analysis tools (Science Tools version 10-00-02)
available at the FSSC~\footnote{\corr{\url{http://fermi.gsfc.nasa.gov/ssc/data/analysis/software}}}.
The maximum likelihood (ML) method implemented in the {\it gtlike} tool can be applied in two different ways, either on a
photon basis (``unbinned ML'' hereafter) or binning the data in energy and sky position (``binned ML'').
In this so-called ``forward-folding'' spectral reconstruction method, the LAT effective area and point spread function are
folded with a source model to compute the number of predicted counts in the RoI (or the photon density for the unbinned
case).
The model includes the spectrum of GRB~090926A and of the background, whose parameters are fitted by comparing the
expected and observed numbers through the maximization of the likelihood function.
The background in the transient class events selected in Sect.~\ref{subsec:data} is mainly composed of charged cosmic rays
which have been misclassified as gamma rays.
It includes also astrophysical gamma rays coming from galactic and extragalactic diffuse and point sources.
In the case of GRB~090926A, the galactic emission could be neglected due to its high galactic latitude ($b=-49.4^\circ$). 
For these reasons, we simply used a power-law to describe the spectrum of the background, with an amplitude
and a spectral index left free to vary.
The spectrum of the GRB was fitted using either a power-law or adding a spectral cutoff at high energy
(see~Sect.~\ref{subsec:models}).\\

For the binned ML case, the {\it gtlike} tool offers the possibility to account for energy dispersion, at the cost of a
slight increase in computing time.
This allowed us to extend our analyses to Pass~8 events with energies below 100~MeV, i.e. to an energy domain where the
LAT energy redistribution function is the widest and can affect the spectral reconstruction if not taken into account.
Therefore, all of our analyses which include Pass~8 data below 100~MeV were performed with the binned ML method and
correcting for the energy dispersion effect.
As reported in Sect.~\ref{sec:results}, spectral analyses above 100~MeV were also performed using the binned and unbinned
versions of the ML method in order to illustrate the gain in LAT sensitivity from Pass~7 to Pass~8 data, and the
consistency between all of these analyses.

\subsection{Joint GBM-LAT spectral analysis}
\label{subsec:joint_ana}
The joint GBM-LAT spectral analyses were performed with the {\it rmfit} tool (version 3.2)
\corr{available at the FSSC~\footnote{\url{http://fermi.gsfc.nasa.gov/ssc/data/analysis/rmfit}}}, using the Castor fit
statistic in order to account for the low counts in the LAT data.
In these analyses, we prepared the LAT data using the aforementioned Science Tools.
We binned the LAT data in energy with the {\it gtbin} tool, and we used the {\it gtbkg} tool to provide {\it rmfit} with a
count spectrum of the background, based on the best model parameters obtained from the fitting procedure described in
Sect.~\ref{subsec:lat_ana}.\\

The count spectrum of the background in the GBM was obtained by fitting background regions of the light curve before and
after the burst, using the same time intervals as in~\cite{090926a}.
In addition, we followed the methodology described in~\cite{090926a} regarding the global effective area correction
to be applied to the BGO data, due to the relative uncertainties in the NaI and BGO detectors responses.
In order to match the flux given by the NaI detectors, a normalization factor $f_\mathrm{eff}$ between the two types of
detectors (NaI and BGO) was introduced in the fit.
We left $f_\mathrm{eff}$ free to vary and we estimated it by fitting the whole prompt emission spectrum (i.e. from
T$_{0}$ + 3.3~s to T$_{0}$ + 21.6~s).
The fitted value $f_\mathrm{eff}=0.825\pm0.013$ is marginaly compatible with the value of $0.79$ reported
in~\cite{090926a}. We also checked that this slight difference did not affect our results.
In all of our joint analyses presented in Sect.~\ref{sec:results}, $f_\mathrm{eff}$ was held fixed at $0.83$.

\subsection{Spectral models}
\label{subsec:models}
GRB~090926A was analysed with different spectral models, which are chosen among the functions described below, or
combinations of them.
All functions are normalized by a free amplitude parameter $A$, in units of cm$^{-2}$\,s$^{-1}$\,keV$^{-1}$.
Following~\cite{090926a}, the spectra are always represented by the phenomenological Band function~\citep{band93}
in the keV-MeV domain. This function is composed of two smoothly-connected power laws with 4 free parameters
($\corr{A_B}$, $E_p$, $\alpha$ and $\beta$):
\lupm{
\begin{equation*}
\frac{dN}{dE}\big(E\,|\,A_B,E_p,\alpha,\beta\big) = 
\end{equation*}
\begin{equation}
  A_B \left\{
  \begin{array}{l l}
   \left(\frac{E}{100\,\mathrm{keV}}\right)^{\alpha} \exp\left(-\frac{E\,(2+\alpha)}{E_p}\right), &  E \le E_p {{\alpha-\beta}\over {2+\alpha}} \\
   \\
   \left(\frac{E}{100\,\mathrm{keV}}\right)^\beta \left(\frac{E_p}{100\,\mathrm{keV}}\frac{\alpha-\beta}{2+\alpha}\right)^{\alpha-\beta}
   \exp\left(\beta-\alpha\right) , & E > E_p {{\alpha-\beta}\over {2+\alpha}}
\end{array}
\right.
\label{eq:band}
\end{equation}
}
where $\alpha$ and $\beta$ are the respective photon indices, and $E_p$ is the peak energy of the spectral energy
distribution (SED), $\displaystyle \nu F_\nu=E^2\frac{dN}{dE}$.\\

In the LAT energy range, we adopted either a power-law (hereafter PL), a power-law with exponential
cutoff (CUTPL), or a broken power law with exponential cutoff (CUTBPL).
The CUTBPL function has 3 free parameters ($\corr{A_C}$, $\gamma$ and $E_f$) and is defined as: 
\lupm{
  \begin{equation}
\frac{dN}{dE}\big(E\,|\,A_C,\gamma,E_f) = A_C \left\{
\begin{array}{l l}
   \left(\frac{E}{E_{piv}}\right)^{\gamma_0}\exp\left(-\frac{E}{E_f}\right), &  E \le E_b \\
\\
    \left(\frac{E_b}{E_{piv}}\right)^{\gamma_0}\left(\frac{E}{E_b}\right)^\gamma \exp\left(-\frac{E}{E_f}\right), & E > E_b
\end{array}
\right.
\label{eq:cutbpl}
\end{equation}
}
where $\gamma$ is the photon index and $E_{f}$ is the folding energy \heb{of the exponential cutoff} that characterizes
the high-energy spectral break.
At low energies, the break at $E_b=200$~keV and the photon spectral index $\gamma_0=+4$ have been fixed
in order to ensure that the flux in the keV-MeV domain is negligible with respect to the flux from the Band
spectral component, as expected from an emission spectrum consisting of a synchrotron component in the keV-MeV domain and
an inverse Compton component at higher energies (see Sect.~\ref{subsec:joint_res}).
\lupm{Specifically, the break energy $E_b$ has been fixed to the value that is obtained when this parameter is left free to
  vary.
  In order to minimize the correlation between the fitted parameters, the pivot energy $E_{piv}$ has been chosen close
  to the decorrelation energy. It has been fixed to a value between 200~MeV and 500~MeV in the LAT-only spectral analyses, and to 10~MeV
  (time interval \cc) or 100~MeV (time interval \dd) in the GBM-LAT joint spectral fits.}\\

The general formulation in Eq.~(\ref{eq:cutbpl}) defines the PL and CUTPL functions as subsets of the CUTBPL function.
The CUTPL function is obtained in the limit $E_b \rightarrow 0$ and
redefining the amplitude parameter as $\displaystyle \corr{A_C}\rightarrow A'=\corr{A_C}\left(\frac{E_b}{E_{piv}}\right)^{\gamma_0}$.
It has 3 free parameters ($A'$, $\gamma$ and $E_f$):
\begin{equation}
\frac{dN}{dE}\big(E\,|\,A',\gamma,E_f) = A' \left(\frac{E}{E_{piv}}\right)^\gamma \exp\left(-\frac{E}{E_f}\right).
\label{eq:cutpl}
\end{equation}
The PL function is obtained by further imposing $E_f \rightarrow +\infty$ (1~TeV in practice), leaving 2 free parameters
($A'$ and $\gamma$): 
\begin{equation}
\frac{dN}{dE}\big(E\,|\,A',\gamma) = A' \left(\frac{E}{E_{piv}}\right)^\gamma.
\label{eq:pl}
\end{equation}

In the analyses presented in Sect.~\ref{sec:results}, we estimated the significance of the high-energy spectral break by
fitting models with and without an exponential cutoff at the highest energies.
For the LAT-only spectral analysis (Sect.~\ref{subsec:lat_ana}), we computed the test statistic
$\displaystyle TS=2\,(\ln\mathcal{L}_{1} - \ln\mathcal{L}_{0})$, where $\mathcal{L}_{0}$ and $\mathcal{L}_{1}$ are the
maximum values of the likelihood functions obtained with the PL and CUTPL models, respectively.
For the joint GBM-LAT spectral analysis (Sect.~\ref{subsec:joint_ana}), $TS$ is simply given by the decrease in Castor
fit statistic $\Delta C_\mathrm{stat}$ when an exponential cutoff (i.e. the $E_f$ parameter) is added to the high-energy
power-law component of the spectral model.
In the large sample limit, $TS$ is equal to the square of the spectral break significance, thus we approximated the
latter as $\displaystyle N_\sigma\simeq\sqrt{TS}$.


\section{Results}
\label{sec:results}
This section presents the results of our spectral analyses.
More information on the spectral fits are given in the tables in the Appendix (Sect.~\ref{sec:appendix}).
Firstly, we performed a spectral analysis of GRB~090926A using LAT-only data over the burst duration at high energy and
focusing on the time interval \cc (Sect.~\ref{subsec:lat_res}).
Then, we performed a time-resolved spectral analysis of GRB~090926A through joint fits to the GBM and LAT data during the
time intervals \cc and \dd, revealing the time evolution of the high-energy spectral
break (Sect.~\ref{subsubsec:best_model} and~\ref{subsubsec:time_evolution}). 
\eng{We carefully studied} the stability of these results with respect to the systematic uncertainty on the LAT response
(Sect.~\ref{subsubsec:systematics}). 
Finally, we estimated the variability time scales in time intervals \cc and \dd and for the GBM and LAT energy ranges
(Sect.~\ref{subsec:variability}), which are needed for the theoretical interpretation presented in the next section.

\subsection{LAT-only spectral analysis}
\label{subsec:lat_res}
\begin{figure*}[!t]
  \centering
  \includegraphics[width=0.7\linewidth]{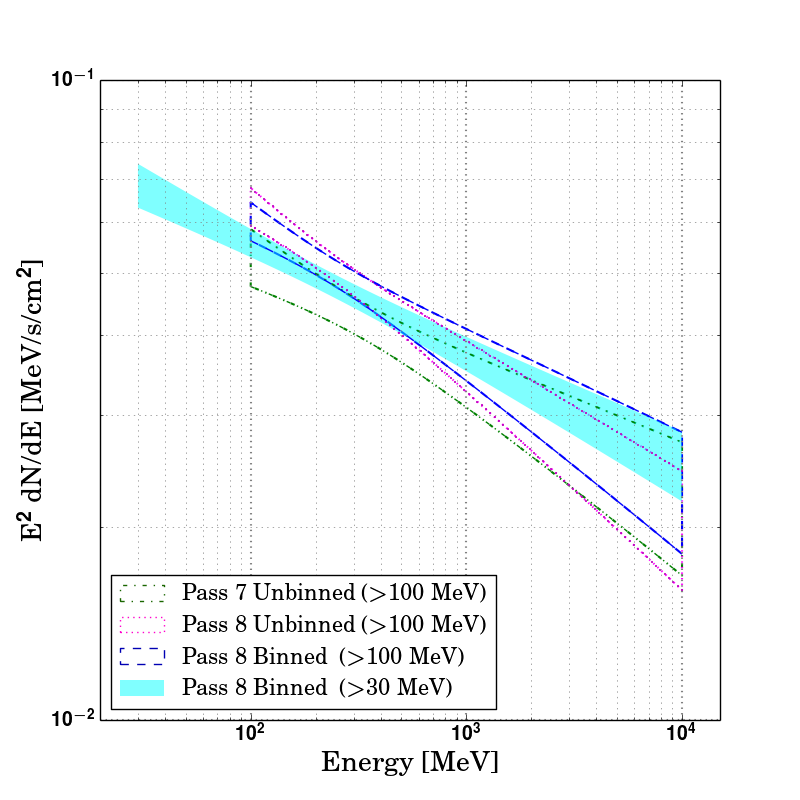}
  \caption{
    GRB~090926A time-averaged spectral energy distribution as measured by the \Fermi/LAT, using Pass~7 data above 100~MeV
    \corr{(dotted-dashed butterfly)} and Pass~8 data above 30~MeV \corr{(filled butterfly)} or 100~MeV \corr{(dotted and
    dashed butterflies)}. 
    Each spectrum is represented by a 68\% confidence level contour \corr{derived from the errors on the parameters of the
      fitted power-law function.}
  }
  \label{fig:sed_lat}
\end{figure*}
\eng{We first characterized} the time-averaged spectrum of GRB~090926A, using LAT Pass~7 and Pass~8
data above 100~MeV during the \latdur time interval.
We fitted the spectrum with a PL model using the unbinned ML method with Pass~7 data, and the unbinned and binned ML methods
with Pass~8 data.
As shown in Table~\ref{tab:fit_latt90_pl} in the Appendix, all of the three fits gave consistent results in terms of
photon index and integrated flux above 100~MeV.
Using Pass~8 data yielded spectral parameters that are slightly more constrained.
A better accuracy was reached by applying the binned ML analysis to Pass~8 data including events with energies down to
30~MeV (see the last column of Table~\ref{tab:fit_latt90_pl}).
The number of events was $\sim$$2.3$ times larger in this configuration, yielding a photon index $\gamma=-2.20\pm0.03$ and an
integrated flux of $(48\pm 1.5)\times10^{-5}$cm$^{-2}$\,s$^{-1}$.
Fig.~\ref{fig:sed_lat} shows the SED of GRB~090926A for the four analyses described above, and their excellent agreement.
The narrowest confidence level contour, \corr{shown as a filled butterfly in the figure,} is obtained for the last fit,
which clearly illustrates the improvement of the spectral reconstruction with Pass~8 data above 30~MeV.\\

\eng{Then,} we focused our analysis on the time interval \cc, where the high-energy spectral break of GRB~090926A
was initially found~\citep{090926a}.
We fitted the spectrum with PL and CUTPL models using the same ML methods and data sets as for the analyses of the \latdur
time interval described above.
As shown in Table~\ref{tab:fit_lat_c_cutpl} in the Appendix, the precision on the fitted photon index is poor due to the
low event statistics above 100~MeV. Moreover, no significant spectral break was found in these analyses.
However, the binned ML analysis of Pass~8 data above 30~MeV yields a marginal detection ($N_\sigma=4.4$), with a
\heb{folding} energy $E_f=0.41_{-0.14}^{+0.27}$~GeV and a photon index $\gamma=-1.68\pm0.22$.
These values are affected by large errors, and they are fully compatible with the more accurate measurements reported
in~\cite{090926a}, which \eng{were} obtained using GBM and LAT data in a joint spectral fit.

\subsection{Joint GBM-LAT spectral analysis}
\label{subsec:joint_res}
\subsubsection{Spectrum representation}
\label{subsubsec:best_model}
The results presented in Sect.~\ref{subsec:lat_res} indicate that the high-energy spectral break of GRB~090926A is hard to
detect with LAT-only data.
Therefore, we considered GBM and LAT data in a joint spectral fit in order to bring in additional constraints on the photon
index of the high-energy component, and to increase the sensitivity to any possible spectral break.
Starting with a single Band component in the spectral model, we reanalyzed the three time intervals \bb, \cc and \dd using
GBM and LAT Pass~7 or Pass~8 data.
Then, we increased the complexity of the model by adding an extra high-energy PL component, or a CUTPL component to
search for the presence of a spectral break.
Similarly to~\cite{090926a}, the extra PL component was found to be very significant in the time intervals \cc and
\dd only.
Nevertheless, we searched for a possible spectral break in the time interval \bb, i.e. we added an exponential attenuation
to the high-energy power-law branch of the Band component.
We found that a spectral break is not required by the data, and that the Band model is enough to reproduce the
spectrum of GRB~090926A in this time interval.\\

In time interval \cc, the comparison of the Band+PL and Band+CUTPL joint fits confirmed the presence of a high-energy
spectral break in the extra PL.
Not surprisingly, the evidence for this break was found to be the highest using LAT Pass~8 data above 30~MeV in the spectral
fit, i.e. the best LAT data set with the widest spectral coverage.
As shown in Table~\ref{tab:fit_joint_c_cutpl} and Fig.~\ref{fig:sed_band+cutpl} \corr{(top panel)} in the Appendix, the break
significance increased from $N_\sigma=5.9$ with Pass~7 data above 100~MeV to $N_\sigma=7.7$ with Pass~8 data above 30~MeV.
Moreover, both the photon index $\gamma=-1.68_{-0.03}^{+0.04}$ and the \heb{folding} energy $E_f=0.37_{-0.05}^{+0.06}$~GeV of the
CUTPL component were very well constrained in the latter case. \eng{These} values are compatible with the results reported
in~\cite{090926a}, $\gamma=-1.71_{-0.05}^{+0.02}$ and $E_f=0.40_{-0.06}^{+0.13}$~GeV.\\

In time interval \dd, a marginal detection of a spectral break ($N_\sigma\sim4$) was reported in~\cite{090926a}, \corr{with $E_f=2.2_{-0.7}^{+0.9}$~GeV}.
As shown in Table~\ref{tab:fit_joint_d_cutpl} and in Fig.~\ref{fig:sed_band+cutpl} \corr{(bottom panel)} in the Appendix, we found
a similar significance of $N_\sigma=4.3$ when using LAT Pass~7 data above 100~MeV in the joint spectral fit.
Using instead Pass~8 data above 30~MeV, the significance increased to $N_\sigma=5.8$.
This Band+CUTPL fit yielded a photon index $\gamma=-1.75_{-0.03}^{+0.02}$ which is \corr{similar to the one found in} the time
interval \cc, and a \heb{folding} energy $E_f=1.61_{-0.31}^{+0.38}$~GeV which is significantly higher.
These results reveal for the first time that the spectral attenuation persists at later times, with an
increase of the break \heb{characteristic} energy up to the GeV domain, and until the end of the \corr{keV-MeV} prompt emission phase of
GRB~090926A, \corr{as measured by the GBM}.\\
\begin{table*}[!t]
  \caption{
    Results of the Band+CUTBPL fits to GBM and LAT data during the time intervals \cc, \dd, \dda and \ddb.
    \lupm{The pivot energy $E_{piv}$ in equation (\ref{eq:cutbpl}) has been chosen close to the decorrelation energy. It
      has been fixed to 10~MeV for interval \cc and 100~MeV for intervals \dd, \dda and \ddb.}
    The last row shows the increase in $C_\mathrm{stat}$ with respect to fits with a Band+CUTPL model.
    \corr{In order to facilitate the comparison with the results from the LAT-only spectral analysis ,
      we also indicated the values of the photon index $\gamma$ and of the \heb{folding} energy $E_f$ found in
      Sect.~\ref{subsec:lat_res} for the time interval \cc.}
  }                    
  \label{tab:fit_joint_cutbpl}
  \centering
  \begin{tabular}{l c c c c c}
    \hline
    \hline
    Time interval &  \cc & \dd  & \dda & \ddb \\
    \hline
    Time interval boundaries from T$_{0}$~(s)&9.8-10.5 & 10.5-21.6& 10.5-12.9&12.9-21.6\\
    \lupm{Band amplitude $A_{B}$ ($\times$10$^{-2}$ cm$^{-2}$\,s$^{-1}$\,keV$^{-1}$) }& \lupm{$34_{-1}^{+2}$} & \lupm{$10.1_{-0.3}^{+0.2}$}  &\lupm{$29_{-1}^{+1}$}  &\lupm{$4.7_{-0.2}^{+0.1}$}\\
    Band $E_\mathrm{peak}$ (keV) & $190_{-9}^{+9}$ & $177_{-3}^{+7}$  & $198_{-10}^{+6}$  & $143_{-7}^{+4}$\\
    Band photon index $\alpha$ & $-0.94_{-0.02}^{+0.03}$ & $-0.86_{-0.03}^{+0.01}$ & $-0.73_{-0.04}^{+0.01}$ & $-1.03_{-0.02}^{+0.05}$ \\
    Band photon index $\beta$  & $-3.2_{-0.9}^{+0.2}$ & $-3.1_{-0.5}^{+0.2}$   & $-3.1_{-0.4}^{+0.2}$  & $-3.7_{-1.5}^{+0.3}$ \\
    \lupm{CUTBPL amplitude $A_{C}$ ($\times$10$^4$ cm$^{-2}$\,s$^{-1}$\,keV$^{-1}$)} &\lupm{$4.6_{-0.9}^{+0.9}$}
    &\lupm{$(9.4_{-0.1}^{+0.5})\times 10^3$} & \lupm{$(12_{-4}^{+7})\times 10^3$} &\lupm{$(7_{-1}^{+1})\times 10^3$}\\
    CUTBPL photon index $\gamma$ & $-1.48_{-0.08}^{+0.09}$ \corr{($-1.68\pm0.22$)}  & $-1.71_{-0.05}^{+0.05}$ & $-1.55_{-0.10}^{+0.12}$ & $-1.68_{-0.05}^{+0.05}$ \\
    CUTBPL \heb{folding} energy $E_f$ (GeV) & $0.34_{-0.05}^{+0.07}$ \corr{($0.41_{-0.14}^{+0.27}$)}  &  $1.20_{-0.18}^{+0.22}$ & $0.55_{-0.10}^{+0.13}$ & $1.43_{-0.25}^{+0.49}$\\
    Break significance $N_\sigma$ &  7.6 &  6.1 &  4.3 & 5.1 \\
    $C_\mathrm{stat}$ / dof &  604.7 / 518 & 652.7 / 518 &  559.0 / 518 & 603.2 / 518 \\
    $\Delta C_\mathrm{stat}$ &  15.9 & 12.1  & 6.1 & 15.1 \\
    \hline
  \end{tabular}
\end{table*}

In Sect.~\ref{sec:theory}, we will discuss our results in terms of keV-MeV synchroton radiation of electrons
accelerated during the dissipation of the \eng{jet's internal energy}, and inverse Compton emission at higher energies.
In this theoretical framework, the inverse Compton component is not expected to contribute significantly to the flux
at the lowest energies.
The Band+CUTPL representation of GRB~090926A spectra, that we used in the aforementioned analyses, does not meet this
requirement, since the extrapolation of the CUTPL component down to $\sim$10~keV yields a flux that is comparable to the
flux of the Band component~\citep[see Fig.~\ref{fig:sed_band+cutpl} in the Appendix and Fig.~5 in][]{090926a}.
Conversely, the CUTBPL component (see Sect.~\ref{subsec:models}) is more physically motived.
For these reasons, we repeated the joint spectral fits in the time intervals \cc and \dd, using LAT Pass~8 data
above 30~MeV and adopting the Band+CUTBPL model.\\

As shown in the last two rows of Table~\ref{tab:fit_joint_cutbpl}, the choice of the Band+CUTBPL model as the best
representation of GRB~090926A spectra is justified by its ability to reproduce the data adequately in the time intervals
\cc and \dd.
We investigated different representations of the spectral break\corr{, e.g. trying to reproduce the more complex shape
  predicted in Fig.~6 of \cite{hascoet12}, which consists of a broken power law with an exponential attenuation at higher
  energies. However, and similarly to the analysis reported in~\cite{090926a},} the limited
photon statistics prevented us from characterizing the shape of this spectral attenuation better than with the Band+CUTBPL model.
The Castor fit statistics obtained with the Band+CUTBPL model are only slightly larger that the ones obtained with the
Band+CUTPL model ($\Delta C_\mathrm{stat}=15.9$ and $12.1$ for time intervals \cc and \dd, respectively).
In addition, the spectral parameters remained essentially unchanged, the main difference being observed for the Band
photon index $\alpha$, as expected from the different contributions of the CUTPL and CUTBPL components to the low-energy
flux.
This parameter decreased from $\alpha\sim-0.6$ (Band+CUTPL model, \corr{see Tables~\ref{tab:fit_joint_c_cutpl}
  and~\ref{tab:fit_joint_d_cutpl}} in the Appendix) to $\alpha\sim-0.9$ (Band+CUBTPL model).
Both values are larger than the theoretical prediction $\alpha=-3/2$ for pure fast cooling synchrotron~\citep{sari98},
whereas this regime is required to explain the high temporal variability and to reach a high radiative efficiency compatible
with the huge observed luminosities. 
The value $\alpha\sim -0.6$ is difficult to reconcile with synchrotron radiation, except by invoking the marginally fast
cooling regime~\citep{daigne11,beniamini13}.
The value $\alpha\sim -0.9$ found in the Band+CUTBPL model is in better agreement, as it is 
\corr{well below the ``synchrotron death line'' $\alpha=-2/3$ and very}
close to the limit $\alpha\sim
-1$ expected in the fast cooling regime affected by inverse Compton scatterings in the Klein Nishina regime~\citep{daigne11}.
At high energy, the CUTBPL component is slightly harder than the CUTPL component, with a fitted photon index
$\gamma=-1.48_{-0.08}^{+0.09}$ (resp. $-1.71_{-0.05}^{+0.05}$) in the time interval \cc (resp. \dd), whereas the
\heb{folding} energy $E_f=0.34_{-0.05}^{+0.07}$~GeV (resp. $1.20_{-0.18}^{+0.22}$~GeV) and its significance $N_\sigma=7.6$
(resp. $6.1$) are close to the ones previously obtained from the Band+CUTPL fit to the data.

\subsubsection{Time evolution of the high-energy spectral break}
\label{subsubsec:time_evolution}
\begin{figure*}[!t]
  \centering
  \includegraphics[width=\linewidth]{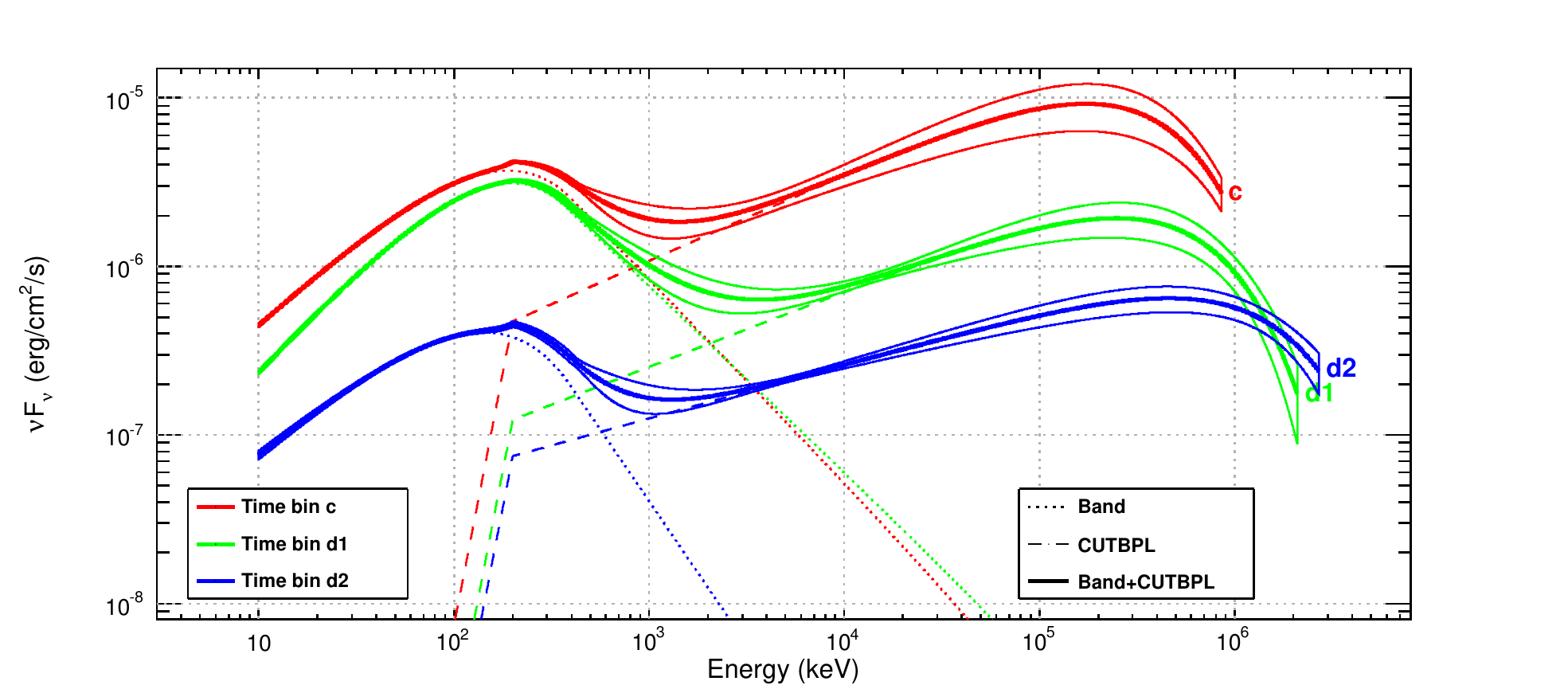}
  \caption{
    GRB~090926A spectral energy distributions as measured by the \Fermi GBM and LAT in time intervals \cc (red curves),
    \dda (green curves) and \ddb (blue curves), using LAT Pass~8 data above 30~MeV.
    \corr{Each solid curve represents the best fitted spectral shape (Band+CUTBPL), within a 68\% confidence level contour
    derived from the errors on the fit parameters.}
  }
  \label{fig:sed_band+cutbpl}
\end{figure*}
The time evolution of the spectral break \heb{characteristic} energy in the extra power-law component of GRB~090926A is a novel result that has
been made possible thanks to the \eng{improved} event statistics in the LAT Pass~8 data set.
We further investigated this spectral evolution by splitting the time intervals \cc and \dd, either dividing them in two
sub-intervals of equal statistics, or isolating the rising and decaying parts in the corresponding light curves.
Then, we performed a Band+CUTBPL using the same procedure as in Sect.~\ref{subsubsec:best_model}.\\

The results of these four fits are reported in Tables~\ref{tab:fit_joint_c_split} and~\ref{tab:fit_joint_d_split} in the
Appendix.
No time evolution was found within the interval \cc, in particular between the two sub-intervals of equal statistics, in
which the high-energy spectral break was detected with high significance ($N_\sigma\geq 5$).
Conversely, the high-energy spectral break was found to evolve within the time interval \dd, with a significance between
$4.1$ and $5.3$ depending on the splitting method.
In the following, we retained the pair of sub-intervals with equal statistics, \dda (from
10.5~s to 12.9~s post-trigger) and \ddb (from 12.9~s to 21.6~s post-trigger).
The results of the Band+CUTBPL fits to GBM and LAT data during these time intervals are summarized in
Table~\ref{tab:fit_joint_cutbpl}.
As for the time intervals \cc and \dd, the Band+CUTBPL model was found to reproduce the data adequately.
Between \dda and \ddb, the \heb{folding} energy $E_f$ increased from $0.55_{-0.10}^{+0.13}$~GeV (with a significance
$N_\sigma=4.3$) to $1.43_{-0.25}^{+0.49}$~GeV ($N_\sigma=5.1$).
The final spectral energy distributions for the time intervals \cc, \dda and \ddb, are represented in
Fig.~\ref{fig:sed_band+cutbpl}, where the increase of the high-energy spectral break from 0.34~GeV (interval \cc) to
1.43~GeV (interval \ddb) is clearly visible.

\subsubsection{Systematic effects}
\label{subsubsec:systematics}
\begin{figure*}[!t]
  \centering
  \hbox{
    \includegraphics[width=0.5\linewidth]{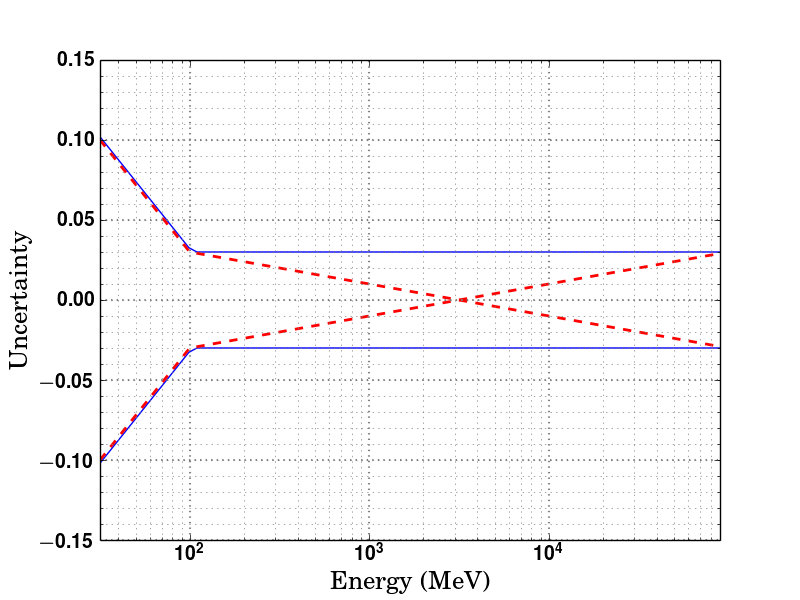}
    \includegraphics[width=0.5\linewidth]{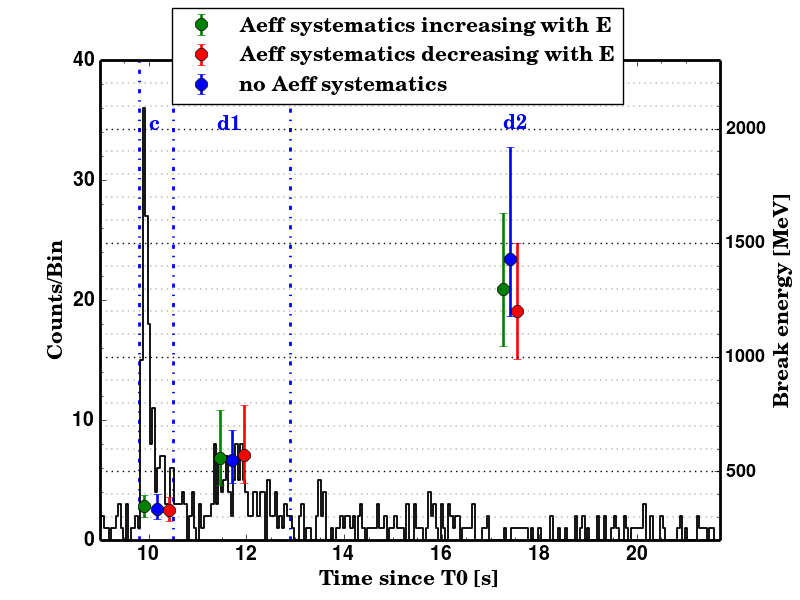}
  }
  \caption{
    Left: Containment interval of the relative systematic uncertainty on the LAT effective area (blue
    curves) as a function of the photon energy $E$, and the two $\epsilon$(E) functions used to estimate the corresponding
    distortion effect on our spectral analysis (red curves); 
    Right: \heb{folding} energies for the time intervals \cc, \dda and \ddb, obtained with or without considering the systematic
    uncertainties on the LAT effective area. \corr{The results have been superimposed to the LAT counts light curve above 30~MeV.}
  }       
  \label{fig:syst}
\end{figure*}

The measurements of the high-energy spectral break of GRB~090926A can be affected by systematic uncertainties due to the
incomplete knowlege of the LAT instrument response functions (IRFs), namely its effective area, point spread function and
energy redistribution function.
\corr{As explained in the FSSC documentation~\footnote{\url{http://fermi.gsfc.nasa.gov/ssc/data/analysis/LAT_caveats.html}},
  the LAT collaboration} has estimated the precision of the instrument simulation by performing several consistency checks
between IRF predictions and data taken from bright gamma-ray sources (the Vela pulsar, bright Active Galactic Nuclei, and
the Earth's limb).
The systematic uncertainty on the effective area is dominant for spectral analyses which account for energy dispersion,
especially below 100~MeV.
The maximum amplitude of this systematic effect has been parameterized as a function of the photon energy
$E$, as represented by the blue curves in the left panel of Fig.~\ref{fig:syst}.
Therefore, we assessed the impact of the systematic effect on the effective area $A_\mathrm{eff}(E)$ by replacing it with
$A_\mathrm{eff}(E)\,[1+\epsilon(E)]$ in the joint spectral fits, where the chosen uncertainty amplitude $\epsilon(E)$ was
constrained within this containment interval.
The two $\epsilon$(E) functions shown in red in the left panel of Fig.~\ref{fig:syst} were found to cause the largest
spectral distortion.
The \heb{folding} energies $E_f$ for the time intervals \cc, \dda and \ddb, obtained with or without twisting the effective area,
are displayed in the right panel of Fig.~\ref{fig:syst}.
As can be seen from this figure, the systematic uncertainty on the LAT effective area does not significantly affect the
results, the observed changes in $E_f$ being negligible with respect to their statistical errors.
In particular, it is worth noting that the confidence intervals on $E_f$ in the different time intervals still exclude
each other after modifying the effective area.


\subsection{Estimation of the variability time scale}
\label{subsec:variability}
The determination of the bulk Lorentz factor of GRB~090926A will be performed in Sect.~\ref{sec:theory} through the
computation of the photon \corr{opacity to pair creation}, which requires a good estimate of the variability time scale,
t$_{v}$, in all time intervals.
For this purpose, we built the light curves for each time interval and in four energy bands, using the summed NaI data
(14~keV-260~keV), the BGO data (260~keV-5~MeV), and the LAT data (30~MeV-10~GeV and 100~MeV-10~GeV).
The first two energy ranges were chosen as in~\cite{090926a}. 
Following~\cite{norris05}, we then fitted each light curve with a temporal profile which is defined as the product of two
exponentials:
\begin{equation}
I(t) = \left\{
\begin{array}{l l}
   0, &  t < t_{s} \\
\\
    A\,e^{-\frac{\tau_{1}}{(t-t_{s})}-\frac{(t-t_{s})}{\tau_{2}}} + B, & t > t_{s}
\end{array}
\right.
\label{eq:pulse}
\end{equation}
where A is a normalization factor, $t_{s}$ is the starting time, $\tau_{1}$ and $\tau_{2}$ are related to the peak time
$\tp=\sqrt{\tau_{1}\tau_{2}}$, and the constant parameter B accounts for the background in each detector.
We used a simple $\chi^{2}$ statistic to check the quality of the fits, and we computed the variability time scale as the
half width at half maximum, \corr{$\displaystyle\tv= \frac{\tau_2}{2}\sqrt{\left(\log(2) + 2\sqrt{\frac{\tau_1}{\tau_2}}\right)^2 - 4 \frac{\tau_1}{\tau_2}}$}.\\

\begin{figure*}[!t]
  \centering
  \hbox{
    \includegraphics[width=0.5\linewidth]{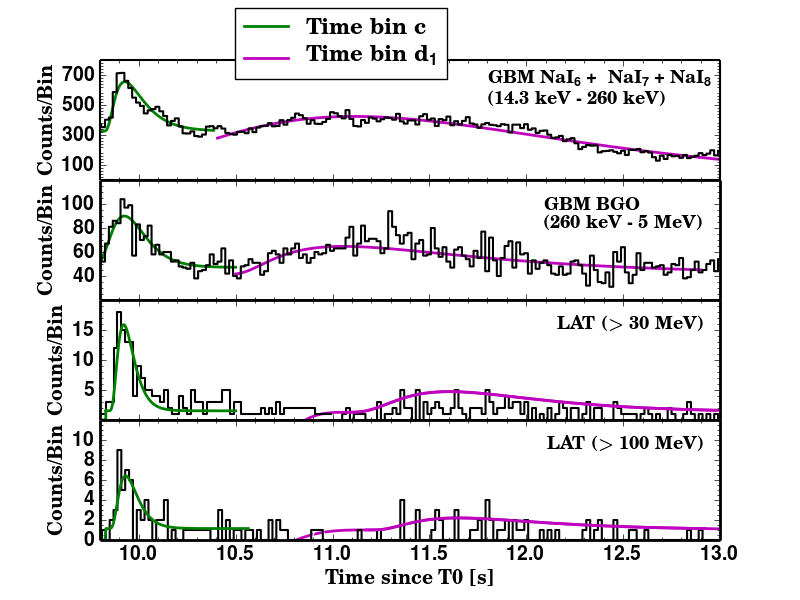}
    \includegraphics[width=0.5\linewidth]{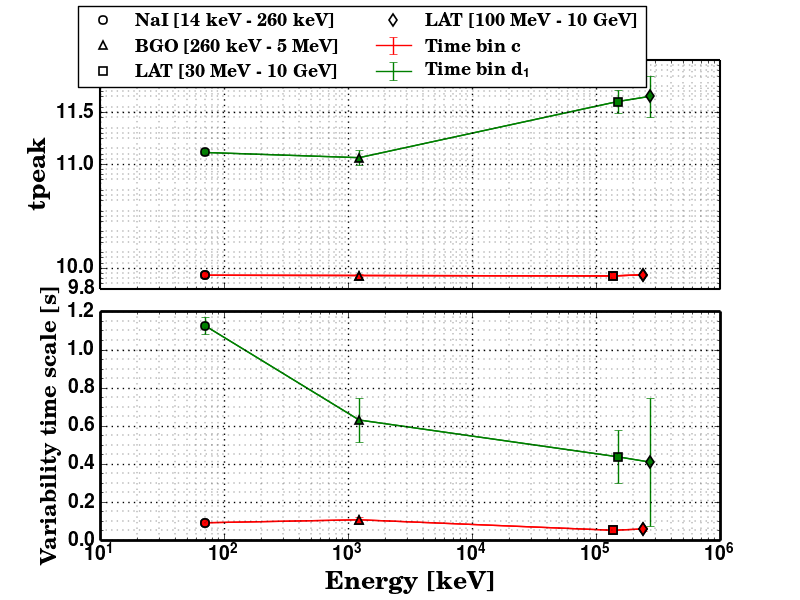}
  }
  \caption{
    Left: GBM and LAT light curves for GRB~090926A with a 0.02~s time binning.
    The data from the GBM NaI and BGO detectors were selected from two energy bands, 14.3–260~keV and 260~keV-5~MeV,
    respectively.
    The last two panels show the LAT light curves above 30~MeV and 100~MeV, respectively;
    Right: Peak time $\tp$ and variability time scale $\tv$ measured for the time intervals \cc (red) and \dda (green) in
    the NaI, BGO and LAT energy bands.
  }
  \label{fig:variability}
\end{figure*}

The left panel of Fig.~\ref{fig:variability} shows the light curves from the beginning of the time interval \cc to
the end of the time interval \dda, along with their temporal fits.
The corresponding values of $\tp$ and $\tv$ are reported in the right panel of the same figure.
In the time interval \cc, our results confirm the remarkable synchronization of the bright spike across the whole
spectrum, with $\tp=9.93\pm0.01$~s in all detectors.
The variability time scales for this time interval is $\tv=0.10\pm0.01$~s in the NaI and BGO energy ranges, and
$\tv=0.06\pm0.01$~s in the two LAT energy ranges.
In the time interval \dda, the time scale decreases from $1.13\pm0.04$~s (NaI) to $0.6\pm0.1$~s (BGO) and $0.4\pm0.2$~s
(LAT).
In both cases, the measured GeV variability time scales thus appear to be similar to the ones in the MeV energy range.
The LAT light curve for the time interval \ddb was not \eng{structured enough} and too difficult to fit.
Conversely, the NaI light curve consists of two different pulses that we fitted with two temporal profiles.
Only the BGO light curve contains a single pulse at $\tp=13\pm0.1$~s, with a variability time scale
$\tv=0.5\pm0.1$~s. 

\section{Interpretation and discussion}
\label{sec:theory}

\subsection{Context}
If the high-energy spectral break observed in the time intervals \cc, \dda and \ddb is an actual cutoff resulting
from pair production $\gamma\gamma\to \mathrm{e^{+}e^{-}}$,
it can be used to estimate the value of the Lorentz factor $\Gamma$ of the
emitting material~\citep{krolik91,baring97,lithwick01,granot08,hascoet12}.
However, the possibility that it could correspond to a natural curvature in the spectrum of the inverse Compton process in
the Klein-Nishina regime (``natural break'' hereafter) cannot be entirely excluded, and only a lower limit on
$\Gamma$ can be obtained in this case.
We consider below these two possibilies in two different scenarios: ({\it i}) the GRB prompt emission in the GBM range is
produced  in the optically thin regime above the photosphere
or ({\it ii}) it is produced at the photosphere, as proposed in the dissipative photosphere
model~\citep{eichler00,rees05,peer06,beloborodov10,beloborodov13}. 
In case ({\it i}), the radius where the  MeV photons are produced is given by
\begin{equation}
R_{\rm MeV}\simeq 2c\,\Gamma^2 \frac{t_{\rm v}}{1+z}\, .
\label{eq:Ris}
\end{equation} 
This estimate corresponds to the internal shock scenario~\citep{rees05,kobayashi97,daigne98} but a comparable radius is
expected in some magnetic reconnection models such as ICMART~\citep{zhang11}.
In both scenarios ({\it i}) and ({\it ii}) the GeV photons can be emitted from the same location as the MeV
photons, $R_\mathrm{GeV}\simeq R_\mathrm{MeV}$, which is expected if the variability of the  GeV and MeV emissions is
comparable (as suggested by our analysis in Sect.~\ref{subsec:variability}), or from a larger radius
$R_\mathrm{GeV}>R_\mathrm{MeV}$ if they come from the further reprocessing of the MeV photons~\citep{beloborodov14} or
have an afterglow origin~\citep{ando08,kumar09,kumar10,ghisellini10,piran10}.

\subsection{Case ({\it i}):  prompt emission produced above the photosphere} 
\label{subsec:abovePh}
\subsubsection{Constraints on the Lorentz factor if the high-energy spectral break is due to gamma-ray opacity to pair creation}
\label{subsubsec:gg}

In this case the radius of the MeV emission is given by Eq.~(\ref{eq:Ris}) above.
If the high-energy spectral break is an actual cutoff resulting from photon opacity to pair creation, the Lorentz
factor can be directly estimated from the burst parameters \citep[Eq.~59 in][]{hascoet12}:
\begin{eqnarray}
\Gamma_{\gamma\gamma}&=& {K\,\Phi(\index)\over \left[{1\over 2}
\left(1+{R_{\rm GeV}\over R_{\rm MeV}}\right)\left({R_{\rm GeV}\over R_{\rm MeV}}\right)\right]^{1/2}}
\ (1+z)^{-(1+\index)/(1-\index)}\cr
& \times&\ \{\sigma_{\rm T}\left[{D_{\rm L}(z)\over c t_{\rm v}}\right]^2 E_\ast F(E_\ast)\}^{1/2(1-\index)}
\ \left[{E_\ast\,E_{\rm cut}\over (m_e c^2)^2}\right]^{(\index+1)/2(\index-1)}.
\label{eq:GG}
\end{eqnarray}

\begin{table*}[!t]
\caption{
   Burst parameters (variability time scale, spectral parameters and luminosity) for the three considered time
     intervals.
     The last four lines give the obtained constraints for the two cases considered in Sect.~\ref{subsec:abovePh}: either the
     observed high-energy spectral break $E_f$ is due to gamma-ray opacity to pair creation, which leads to a
     measurement of the Lorentz factor, or it is a natural break (last line).
     In this second case only a lower limit on the Lorentz factor can be obtained from the maximum energy $E_\mathrm{max}$
     of the observed photons.
     In both cases, the results
     listed in the table corresponds to the assumption $R_{\rm GeV}=R_{\rm MeV}$, favored by the observed similar
     variability at low and high energy. The impact of $R_{\rm GeV}>R_{\rm MeV}$ on these results is illustrated in
     Fig.~\ref{fig:plotGgg}.}
%
\label{tab:paramGG}
\centering 
\begin{tabular}{l c c c c}
\hline
\hline
Time interval & \cc &  \dda & \ddb\\
\hline
$t_{\rm v}$ (s) & $0.1\pm0.01$ & $0.6\pm 0.1$ & $0.5\pm 0.1$ \\
$\index$ & $-1.55^{+0.07}_{-0.09}$ & $-2.25^{+0.10}_{-0.08}$ & $-2.19^{+0.09}_{-0.04}$ \\
$\Phi(\index)$ & $0.511\pm0.009$ & $0.463\pm0.004$ & $0.465\pm 0.003$ \\
$E_f$ (GeV) & $0.34^{+0.07}_{-0.05}$ & $0.55^{+0.13}_{-0.10}$ & $1.43^{+0.49}_{-0.25}$ \\
$E_{\rm max}$ (GeV) & $0.85$ & $2.04$ & $2.66$ \\
$E_\ast$ (MeV) & 10 & 2.5 & 1.0 \\ 
$F(E_\ast)$ (10$^{-2}$ cm$^{-2}$\,MeV$^{-1}$) & $0.22\pm 0.03$ & $4.0\pm 0.8$ & $5\pm 1$  \\ 
Luminosity ($10^{53}$ erg.s$^{-1}$) & $16.9\pm 3.1$ & $1.73\pm 0.14$ & $1.85\pm 0.15$\\
Lorentz factor \iap{$\Gamma_{\gamma\gamma}\left(E_f\right)$} 
& $233\pm 18$ & $100 \pm 8$ & $98\pm 9$ \\ 
Emission radius $R=R_\mathrm{MeV}=R_\mathrm{GeV}$ ($10^{14}\, \mathrm{cm}$) 
& $1.1\pm 0.1$ & $1.2\pm 0.2$ & $1.0\pm 0.1$ \\ 
Photospheric radius $R_\mathrm{ph}$ ($10^{14}\, \mathrm{cm}$) 
& $0.5\pm 0.2$ & $0.7\pm 0.2$ & $0.8\pm 0.2$\\ 
Lower limit on the Lorentz factor \iap{$\Gamma_{\gamma\gamma}\left(E_\mathrm{max}\right)$}
& $257\pm 17$ & $129\pm 8$ & $110\pm 8$ \\ 
\hline
\end{tabular}
\end{table*}  

The various observed quantities appearing in Eq.~(\ref{eq:GG}) are listed in Table~\ref{tab:paramGG} for time intervals
\cc, \dda and \ddb:
$t_{\rm v}$ is the observed variability time scale in the considered time interval, estimated in
Sect.~\ref{subsec:variability};
\iap{$E_\mathrm{cut}$ is the cutoff energy, which we assume here to be equal to the folding energy $E_f$ that characterizes the spectral break
(Sect.~\ref{subsec:joint_res}), also listed in Table~\ref{tab:paramGG};} 
$E_\ast$ is the typical energy of the seed photons interacting 
with those at the cutoff energy $E_{\rm cut}$, $\index$ is the photon index of the seed spectrum close to $E_\ast$ and $F(E_\ast)$ is the photon fluence at $E_\ast$ integrated over a duration $t_{\rm v}$, so that the seed photon spectrum can be approximated by 
$F(E) = F(E_\ast)\left(E/E_\ast\right)^{\,\index}$ (cm$^{-2}$\,MeV$^{-1}$). The energy $E_\ast$  is given by 
\begin{equation}
E_\ast\simeq \frac{\left(2 \Gamma \,m_\mathrm{e}c^2\right)^2}{(1+z)^2 E_\mathrm{cut}}\simeq 1.1\, \mathrm{MeV}\, \left(\frac{\Gamma}{100}\right)^2\left(\frac{E_\mathrm{cut}}{1\, \mathrm{GeV}}\right)^{-1}\, ,
\label{eq:E*}
\end{equation}
$E_\ast$ and $E_\mathrm{cut}$ being the observed values.
As the seed photon spectrum is approximated locally by a power-law, a precise value of $E_\ast$ is not required, as long
as the correct region of the spectrum has been identified. 
\iap{It can be seen that the Lorentz factor $\Gamma_{\gamma\gamma}$ does not depend on a specific choice of $E_\ast$ as
  long as this energy remains in a region where the spectrum keeps a fixed spectral index $s$. Indeed $E_\ast$ appears in two factors in Eq.~(\ref{eq:GG}) with opposite scaling:
$\left(E_\ast F(E_\ast)\right)^{1/2(1-s)}\propto E_\ast^{-(s+1)/2(s-1)}$. }
For the time interval \cc, the seed photons belong clearly to
the CUTBPL component (see Fig.~\ref{fig:sed_band+cutbpl} at $\sim$$10$~MeV),
whereas for the time intervals \dda and \ddb, $E_\ast$ is in the flat transition region of the spectrum  where the
Band and the CUTBPL components overlap (see Fig.~\ref{fig:sed_band+cutbpl}).
In order to quantify the photon index $\index$, we built its distribution using the results of the
  spectral fits.
  Specifically, we assumed that the 7 parameters of the Band+CUTBPL spectral model follow a multi-dimensional gaussian
  distribution. Using their covariance matrix provided by the spectral fit, we generated 1000 sets of values for these
  parameters. For each generated spectrum, we computed numerically the photon index at $E_\ast$. The index $\index$ was
  chosen as the most probable value of the final distribution, and its errors were derived from the
  68\% confidence interval around this value.
The corresponding values of $F(E_\ast)$ in the table are deduced from the spectral fits presented in
Sect.~\ref{subsec:joint_res}; the function $\Phi(\index)$ is defined by
\begin{equation}
\Phi(\index)=\left[2^{1+2\index} \mathcal{I}(\index)\right]^{\frac{1}{2(1-\index)}}\, ,
\label{eq:phi}
\end{equation} 
where $\mathcal{I}(\index)$ depends on $\index$ only and equals \citep{hascoet12}:
\begin{equation}
\mathcal{I}(\index)=\int_0^1\frac{y}{\left(1-y^2\right)^{2+\index}}\, g(\index)\, \mathrm{d}\index\, ,
\end{equation}
with $g(y)=\frac{3}{16}(1-y^2)\left[(3-y^4)\ln{\frac{1+y}{1-y}}-2y(2-y^2)\right]$ coming directly from the dependence of the $\gamma\gamma$ cross section on the energy. \\

Finally, the constant $K$  appearing in front of 
Eq.~(\ref{eq:GG}) has been calibrated by \citet{hascoet12} from a detailed time-dependent calculation of the $\gamma\gamma$ opacity taking into
account a realistic geometry for the radiation field, i.e. 
a time-, space- and direction-dependent photon field in the comoving frame, as expected in an outflow with several relativistically moving emitting zones. This calculation, first carried out analytically by \citet{granot08} and then extended numerically by \citet{hascoet12}, is much more realistic than the standard one zone model, used for instance by \citet{lithwick01}.
\iap{It assumes that the Lorentz factor in the outflow varies between a lowest value $\Gamma_\mathrm{min}$ and a highest value $\kappa\, \Gamma_\mathrm{min}$. If the contrast is of the order of $\kappa\sim 2-5$, }
 the calibration factor 
remains in the interval  
\textbf{$K\sim 0.4 - 0.5$.}
\iap{In such a variable outflow,
the value of $\Gamma_{\gamma\gamma}$ obtained from
Eq.~(\ref{eq:GG}) 
corresponds} to the \iap{lowest} 
Lorentz factor $\Gamma_{\rm min}$ in the outflow~\citep{hascoet12}.\\
%
%

\begin{figure*}[!t]
  \centering
  \begin{tabular}{cc} 
    \includegraphics[width=0.5\linewidth]{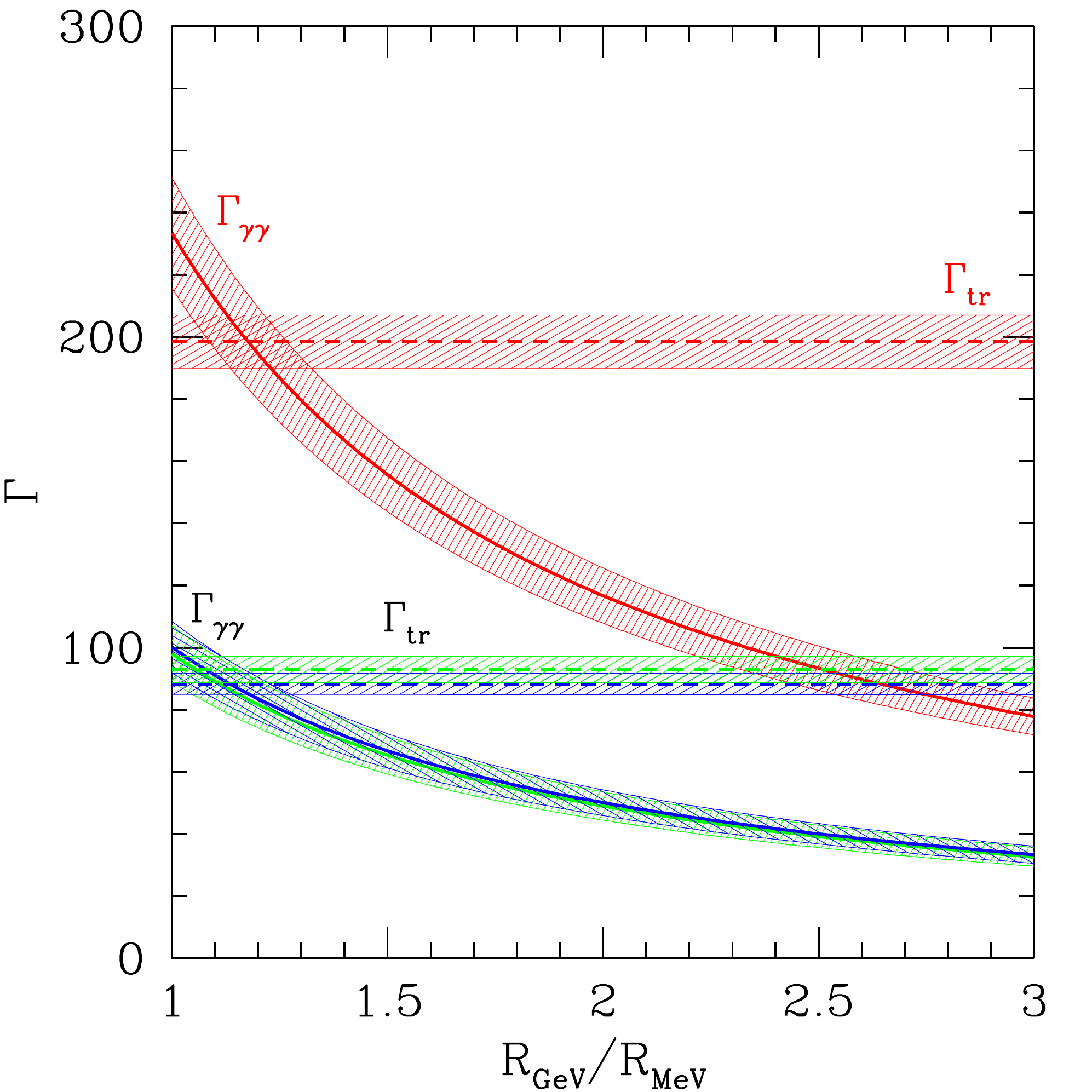} 
    \includegraphics[width=0.5\linewidth]{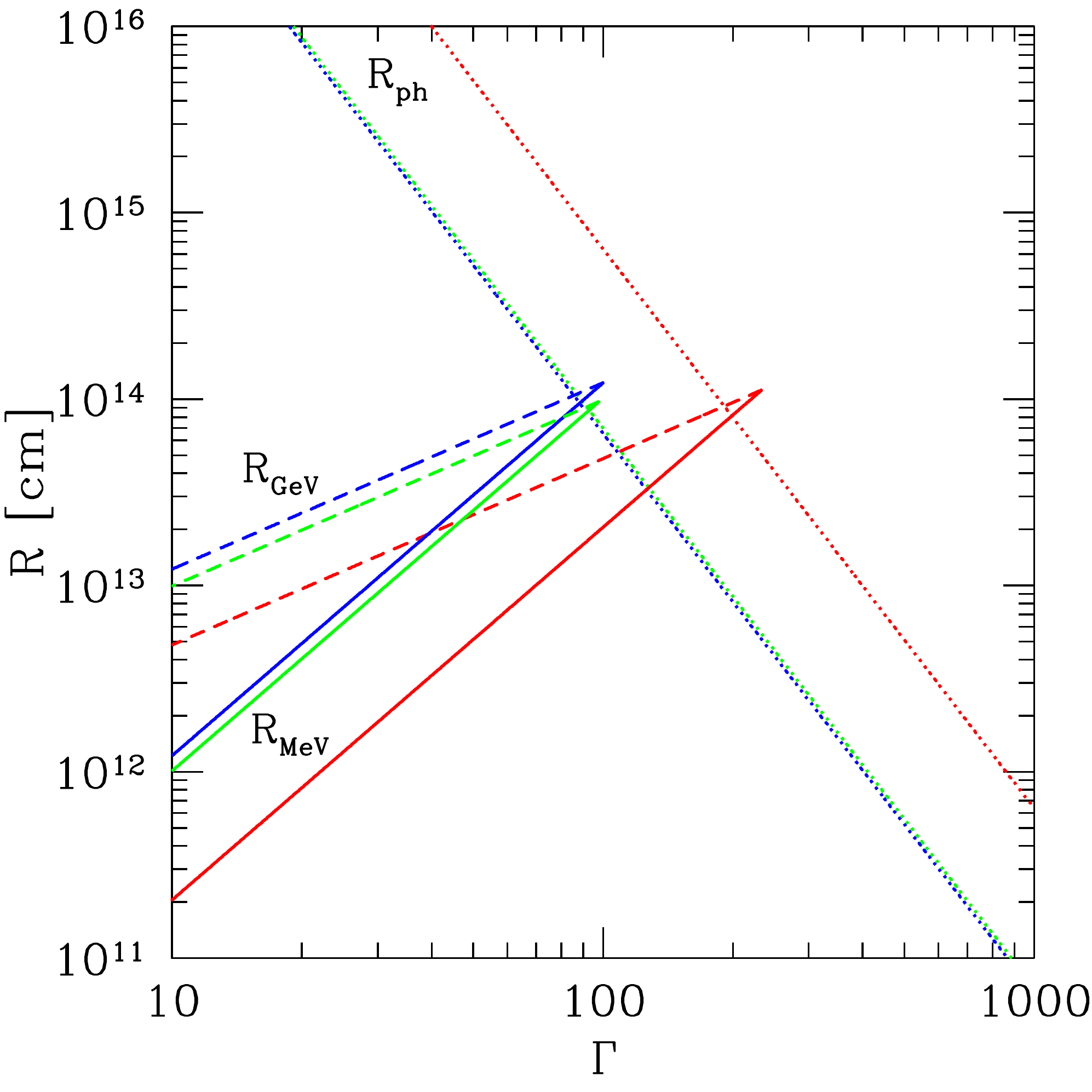}  
  \end{tabular}
  \caption{Left: Lorentz factor $\Gamma_{\gamma\gamma}$ for the time intervals \cc (red), \dda (blue) and \ddb (green) as a function of
    the ratio of the emission radii of the GeV and MeV photons, assuming that the high-energy spectral break comes from
    photon opacity to pair creation \lupm{(Eq.~\ref{eq:GG})}.
    The dashed 
    lines represent the lower limit of the Lorentz factor for transparency, $\Gamma_\mathrm{tr}$
    (Eq.~\ref{eq:Tr}). The shaded strips indicate the typical uncertainty on these quantities, obtained by propagating the errors on the measured values listed in Table~\ref{tab:paramGG}.
    Right: MeV (full lines) and GeV (dashed lines) emission radii as a function of the Lorentz factor. The dotted
    lines correspond to the photospheric radius $R_\mathrm{ph}$ in the different time intervals. 
    The deceleration radius is not plotted, but we checked that it is always well above $R_\mathrm{ph}$, $R_\mathrm{MeV}$ and $R_\mathrm{GeV}$ for normal densities in the external medium (assuming either a wind or a uniform medium).}
  \label{fig:plotGgg}
\end{figure*}

Table~\ref{tab:paramGG} provides the resulting Lorentz factor assuming an equal radius for GeV and MeV emissions. If the
GeV photons are produced at $R_\mathrm{GeV}>R_\mathrm{MeV}$, the Lorentz factor is lower, as can be seen from
Eq.~(\ref{eq:GG}).
The result for each time interval is plotted in Fig.~\ref{fig:plotGgg} (left panel). For
$R_\mathrm{GeV}=R_\mathrm{MeV}$ (as suggested by the comparable variability time scales in the LAT and the MeV
range, see Sect.~\ref{subsec:variability}), we find 
\iap{$\Gamma_{\rm min}=\Gamma_{\gamma\gamma}=233\pm 18$, $100 \pm 8$ and $98\pm 9$}  
for time intervals \cc, \dda and
{\it d$_2$}, respectively. In the time interval \cc, our value is very close to the result of~\cite{090926a}, $\Gamma\simeq 220$, obtained
from a similar analysis based on the detailed analytical approach developed in~\cite{granot08}.
Table~\ref{tab:paramGG} also provides the resulting emission radius $R_\mathrm{MeV}$, which is of the order of $10^{14}$ cm. 
\\

These values for \iap{the lowest Lorentz factor in the outflow} $\Gamma_{\rm min}$ have to be compared with the lower limits on the Lorentz factor for transparency to
Thomson scattering on primary electrons and pair-produced leptons, which corresponds to the assumed condition that the
prompt emission is produced above the photosphere.
This condition reads   
$R_{\rm MeV}\ge R_{\rm ph}$,  with the photospheric radius given by \citep{beloborodov13}
\begin{equation}     
R_{\rm ph}\simeq {\sigma_{\rm T}(1+f\pm)\,{\dot E}\over 8\pi c^3 m_p{\bar \Gamma}^3 (1+\sigma)}\, ,
\label{eq:Rph}
\end{equation}
where ${\bar \Gamma}$ is the average Lorentz factor in the flow, which we approximate by 
\iap{${\bar \Gamma}=\frac{1+\kappa}{2}\Gamma_\mathrm{min}$, $\kappa$ being the contrast defined above}; 
$\sigma_{\rm T}$ is the Thomson cross-section, $f\pm$ the ratio of the number of pairs to primary electrons, ${\dot E}$ 
the total power injected in the flow and $\sigma$ its magnetization at large radius, where the prompt emission is produced, so that
${\dot E}/(1+\sigma)$ is the kinetic power. 
We checked that for the values of the parameters in Table~\ref{tab:paramGG} the optical depth for pair creation is less
than unity at $R_\mathrm{MeV}$. Therefore we adopt  $f\pm =0$ in Eqs.~(\ref{eq:Rph}) and~(\ref{eq:Tr}).
We also assume $\sigma\ll 1$, which is expected for internal shocks. In magnetic reconnection models, if $\sigma$ is
large, $R_{\rm ph}$ is lower and the transparency condition is more easily satisfied. The power $\dot{E}$ is
estimated from the gamma-ray luminosity $L$ listed in Table~\ref{tab:paramGG}  by ${\dot E}=L/\epsilon_{\rm rad}$
assuming a prompt emission efficiency $\epsilon_{\rm rad}=0.1$.
Table~\ref{tab:paramGG} provides the photospheric radius $R_{\rm ph}$ using the measurement of the Lorentz factor
  obtained from the $\gamma\gamma$ constraint.
It can be seen that for $R_\mathrm{GeV}\simeq
R_\mathrm{MeV}$ (as suggested by the comparable variability time scales in the LAT and the MeV range, see
Sect.~\ref{subsec:variability}), the transparency condition is satisfied in all time intervals \cc, \dda and \ddb.
We obtain an emission radius $\sim 10^{14}$~cm and a photospheric radius of a few $10^{13}$~cm in all time intervals.
For $R_\mathrm{MeV}$ given by Eq.~(\ref{eq:Ris}), the transparency condition $R_{\rm MeV}\ge R_{\rm ph}$ yields
\begin{equation}
\bar{\Gamma}>\bar{\Gamma}_{\rm tr}\simeq \left[{\sigma_{\rm T}(1+f\pm)\,{\dot E}\over 8\pi c^4 m_p\,(1+\sigma)\,t_{\rm v}}\right]^{1/5}\ .
\label{eq:Tr}
\end{equation}
The resulting $\Gamma_\mathrm{tr}$ is plotted in
Fig.~\ref{fig:plotGgg} (left panel, horizontal dashed lines).
It appears clearly that the transparency condition can be fulfilled only if $R_\mathrm{GeV}/R_\mathrm{MeV}\le
  1.2-1.3$. As already mentioned, the comparable variability timescales at low and high energy indeed suggest that
  $R_\mathrm{GeV}\simeq R_\mathrm{MeV}$.
When comparing $R_\mathrm{MeV}$ and $R_\mathrm{ph}$, it 
 should be noted that the emission radius deduced from the variability time scale is 
the typical radius where the emission starts. However, the emission continues
at larger radii, as variations on larger time scales are also observed in the light curves.
We conclude from this analysis that GRB 090926A seems fully compatible with the most standard model where the prompt
emission is produced by shocks (or reconnection) above the photosphere.
\\
%

The right panel of Fig.~\ref{fig:plotGgg}, which shows the photospheric and emission (MeV/GeV) radii as a function of the
Lorentz factor, basically contains the same information, presented in a different way.
Again, the figure clearly shows that observations in the three time intervals are compatible with an emission above the
photosphere, as long as the emission radii of the MeV and GeV photons are close to each other.
This is consistent with an internal origin for the high energy component during the prompt phase
suggested by the observed variability.
We stress that this analysis is largely independent of the precise radiative mechanisms.
However, as mentioned in Sect.~\ref{subsubsec:best_model}, a natural candidate is fast cooling synchrotron radiation for
the Band component and inverse Compton scatterings for the CUTBPL component.
Therefore, we discuss below the possibility that the observed spectral break is due to the natural curvature of the latter
component. 

\subsubsection{Constraints on the Lorentz factor if the high-energy spectral break is a natural break}
\begin{figure*}[!t]
  \centering
    \includegraphics[width=0.5\linewidth]{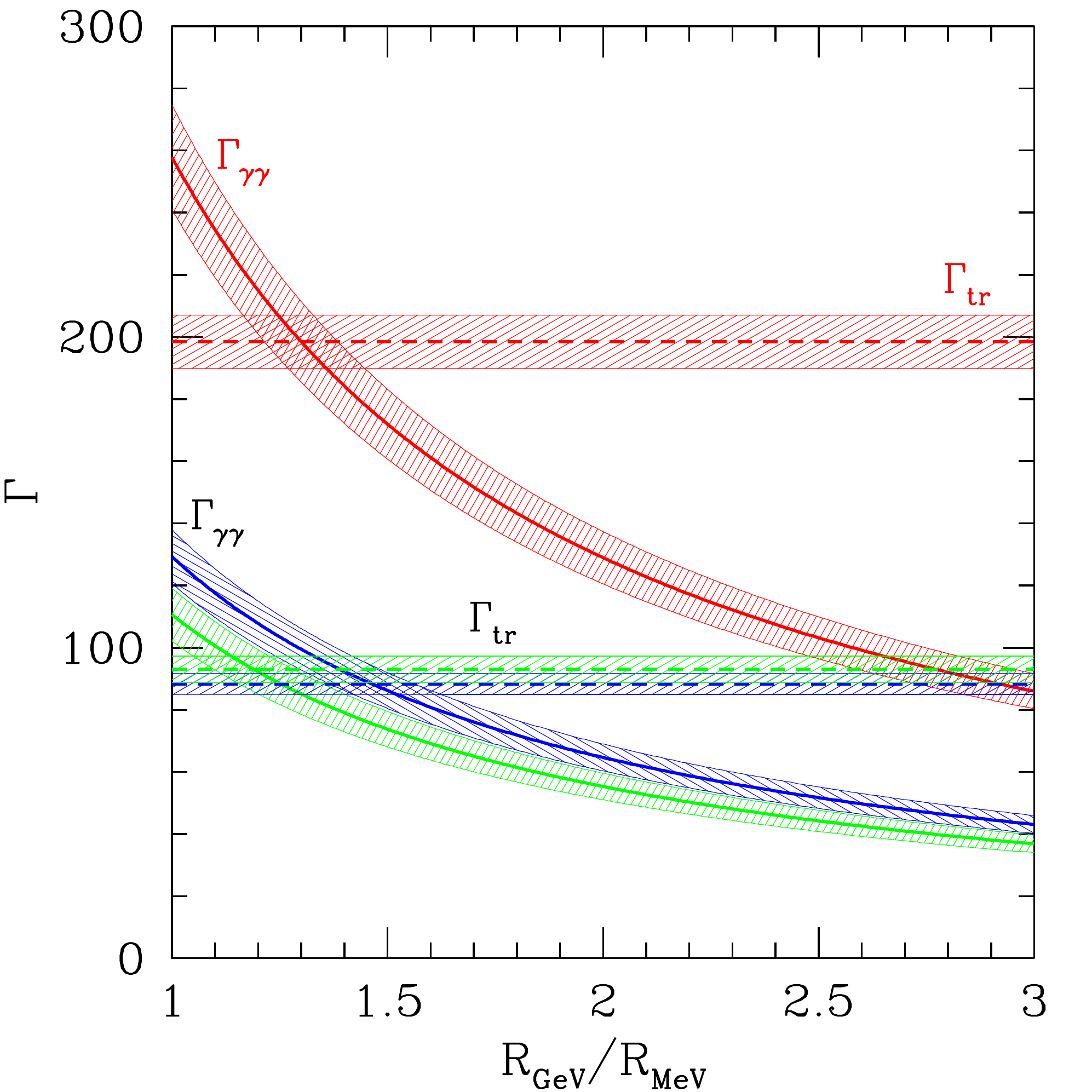} 
  \caption{Same as left panel of Fig.~\ref{fig:plotGgg}, but assuming that the observed cutoff is a natural
      break. The Lorentz factor $\Gamma_{\gamma\gamma}$ is now a lower limit obtained from the photon of highest
      energy in each time interval.}
  \label{fig:plotGmax}
\end{figure*}

If the high-energy spectral break reflects the natural curvature of the inverse Compton spectrum --
namely it does not correspond to photon opacity to pair creation but simply results from the spectral shape of the radiative process --
then only a lower limit on the Lorentz factor can be obtained. It is given, in each time interval, by 
\begin{equation}
\Gamma_{\rm inf}={\rm max}(\Gamma_{\gamma\gamma},\Gamma_{\rm tr})\, ,
\label{eq:Ginf}
\end{equation}
where  $\Gamma_{\gamma\gamma}$ is computed by the same Eq.~(\ref{eq:GG}) as above, using  
the maximum energy $E_{\rm max}$ of the observed photons in 
the time interval (listed in Table~\ref{tab:paramGG}) in place of the \heb{folding} energy $E_f$.
The resulting lower limit on the Lorentz factor is plotted in Fig.~\ref{fig:plotGmax}. 
It shows
that, as soon as $R_{\rm GeV}/R_{\rm MeV}> 1.3$ in the time interval \cc (resp. $1.5$ and $1.2$ in the intervals \dda and
\ddb), the transparency limit (Eq.~\ref{eq:Tr}) becomes more constraining than the limit on the pair-creation opacity.     
However, it has already been mentioned that the variability analysis presented in Sect.~\ref{subsec:variability} rather
suggests $R_\mathrm{GeV}\simeq R_\mathrm{MeV}$. In this case, we find lower limits for the Lorentz factor equal to 
$257\pm 17$, $129\pm 8$ and $110\pm 8$ 
 in time intervals \cc, \dda and \ddb.
\subsection{Case ({\it ii}):  prompt emission produced at the photosphere}
In the case where the prompt emission is produced at the photosphere, 
the radius of the MeV emission is $R_{\rm MeV}=R_{\rm ph}$, given by Eq.~(\ref{eq:Rph}) above.
The constraints derived from the observed spectral break become more difficult to obtain.
Indeed, contrary to the previous case, 
increasing 
the Lorentz factor drives the emitting surface inwards, contributing to increase the optical depth for the GeV photons. 
However pair creation is now expected
below and at the photosphere with values of several tens for $f\pm$ \citep{beloborodov13}, 
which, on the contrary, would contribute to push the photosphere outwards.
Moreover, \eng{since} prompt emission at the photosphere corresponds to lower emission radii $R_\mathrm{MeV}$ than in the optically thin scenario, 
most of the GeV photons could be produced above the photosphere (e.g. at a few $R_{\rm ph}$) 
and still show a short variability time scale ($t_{\rm v}$ value of a few $R_{\rm ph}/2c\Gamma^2$). 
In photospheric models, constraining the Lorentz factor from the high-energy spectral break therefore 
would require a detailed modelling of the radiative transfer from below to above the photosphere, which is beyond
the scope of this paper.
  
\subsection{Discussion}
As shown in the literature, it is widely believed that very bright GRBs should have large Lorentz factors.
  This assumption is based on the pair creation constraint combined with the \Fermi/LAT observations of the first bright
  GRBs (GRBs~080916C, 090510 and 090902B), whose spectrum do not exhibit any attenuation at GeV energies
  \citep{abdo09a,abdo09b,ackermann2010}.
  However, later studies \citep{granot08,hascoet12} pointed out that the single zone model used in these early studies was
  not accurate enough. Large Lorentz factors are also required in models which assume that the GeV emission is produced at
  the external shock, in order to ensure an early deceleration \citep{kumar09,ghisellini10}.
In the present study of GRB 090926A, we disfavor the external origin scenario because of the observed fast variability of the high-energy emission and find values of
the Lorentz factor in the three time intervals reported in Table~\ref{tab:paramGG} 
that are not especially high.
From Eq.~(\ref{eq:GG}), it is interesting to note that a
 lower Lorentz factor and a short variability time scale makes the detection of a cutoff due to pair-creation
   opacity easier since~\citep[see, e.g.,][]{dermer99}
\begin{equation}
E_{\rm cut}\propto \Gamma^{2(\index-1)/(\index+1)}\,t_{\rm v}^{-2/(\index+1)}
\label{eq:Ecut}
\end{equation} 
giving $E_{\rm cut}\propto \Gamma^{10}\,t_{\rm v}^{4}$ and $E_{\rm cut}\propto \Gamma^{5.3}\,t_{\rm v}^{1.7}$ for
$\index=-1.5$ and $-2.2$, respectively.
This could explain why the cutoff in the time interval \cc with $t_{\rm v}=0.1$ s, $\Gamma \sim 230$ and a very large
luminosity  was the most easily accessible. With a larger Lorentz factor the cutoff would have been shifted to a much
higher energy and difficult to characterize.
More generally, the very steep dependence of $E_{\rm cut}$ with the Lorentz factor means that, for a given burst, a cutoff
can be observed in the LAT range for a very small interval in $\Gamma$ only. This may explain why bursts like GRB 090926A are not common in the LAT catalog.\\

As mentioned in Sect.~\ref{subsubsec:gg}, the Lorentz factor that we find in the time \cc is very close to the value
obtained by~\cite{090926a} with the time dependent model of~\cite{granot08}. Another way to
estimate the Lorentz factor consists in assuming that the peak flux time in the LAT or visible range is a good proxy for
the deceleration time of the relativistic ejecta~\citep[see, however,][]{hascoet14}.
With this method~\cite{latgrbcat} found $\Gamma\simeq 600$
(resp. 400) for a uniform  external medium of density $n=1$ cm$^{-3}$ (resp. a stellar wind with a parameter
$A_*=0.1$), assuming a deceleration time  $t_{\rm dec}\simeq10$ s and a gamma-ray efficiency $f_\gamma=0.25$. The obtained
result depends on these parameters as
\begin{equation}
\Gamma\propto \left\lbrace\begin{array}{cl}
& (f_\gamma n)^{-1/8}\,t_{\rm dec}^{-3/8}\ \ \ \ \ {\rm uniform\ external\ medium}\\
& (f_\gamma A_*)^{-1/4}\,t_{\rm dec}^{-1/4}\ \ \ {\rm stellar\ wind}\\
\end{array}\right.
 \label{eq:Gdep}
\end{equation}
which shows that an external medium denser than what was assumed by~\cite{latgrbcat} (e.g. with $n=1000$ cm$^{-3}$
or $A_*=1$) or a deceleration time larger than 10~s could reconcile the two approaches.
A value of $t_{\rm dec}$ larger than 10~s is actually very likely, since the peak flux time in the LAT light curve coincides
with the spike in time interval \cc, which probably results
 from internal dissipation as indicated by its extreme variability.\\

The evolution of the Lorentz factor from the time interval \cc to intervals \dda and \ddb is moderate,
showing a decrease by a factor of 2 while the luminosity in the time interval \cc is 8 times larger and the
variability time scale 5-6 times smaller.
It follows approximately a trend $\Gamma\propto L^{0.3}$ but clearly, with only three time intervals (and two with very
similar temporal and spectral parameters) more data will be needed to check whether this expected behavior~\cite[see, e.g., ][]{baring06} is also found in other bursts and over large parts of their temporal evolution.




\section{Conclusions}
\label{sec:concl}
We presented a new time-resolved analysis of GRB~090926A broad-band spectrum during its prompt phase.
We combined the \Fermi/GBM and LAT data in joint spectral fits in order to characterize the time evolution of its spectrum
from keV to GeV energies, using a Band+CUTBPL spectral model in view of discussing our results in terms of keV-MeV
synchroton radiation of accelerated electrons and inverse Compton emission at higher energies.
In this analysis, we made use of the LAT Pass~8 data publicly released in June 2015, which offer a greater sensitivity
than any LAT data selection used in previous studies of this burst.
Using a Band+CUTBPL model to account for the broad-band spectral energy distribution of GRB~090926A,
we confirmed and better constrained the spectral break at the time of the bright spike, which is observed at $\sim$10~s
post-trigger across the whole spectrum.
Our analysis revealed that the spectral attenuation persists at later times, with an increase of the
break \heb{characteristic} energy until the end of the prompt phase, from 0.34~GeV (interval \cc) to 1.43~GeV (interval \ddb).
We paid careful attention to the systematic effects arising from the uncertainties on the LAT response, and we showed that
this time evolution of the spectral break in the high-energy power-law component of GRB~090926A spectrum is
solid and well established.\\

After computing the variability time scales from keV to GeV energies during and after the bright spike, we discussed our
results in the framework of prompt emission models.
\fpc{We interpret the high-energy spectral break as caused} by photon \corr{opacity to pair creation}.
Requiring that all emissions are produced above the photosphere of GRB~090926A, \fpc{we compute} the bulk Lorentz
  factor of the outflow, $\Gamma$. The latter \corr{decreases from $230$ during the spike to $100$ at the end of the
    prompt emission,} a novel result which improves upon early publications on this burst~\citep{090926a}.
Assuming, instead, that the spectral break reflects the natural curvature of the inverse Compton spectrum,
lower limits corresponding to larger values for $\Gamma$ were also derived.
Despite the increased photon statistics provided in LAT Pass~8 data, we could not favor any of these possible scenarios.
In both scenarios, the extreme temporal variability of GRB~090926A and \fpc{the Lorentz
factors} lead to emission radii \corr{$R\sim10^{14}$~cm and to a photospheric radius of a few $10^{13}$~cm in all time
  intervals.}
This strongly suggests an internal origin of both the keV-MeV and GeV prompt emissions associated \corr{with internal jet
  dissipation} above the photosphere.
This interpretation is reinforced by the flattening of the gamma-ray light curve decay which occurs well after the end of
the keV-MeV prompt emission~\citep{latgrbcat}, \corr{as mentioned in Sect.~\ref{sec:intro}}.\\

In the future, further progress in the understanding of the GRB GeV emission which coincides with the emergence of an
additional power-law component will be possible by using LAT Pass~8 data in broad-band analyses of other LAT bright bursts
with similar temporal and spectral properties \eng{to GRB~090926A,} like the short GRB~090510~\citep{latgrbcat}.
On the theoretical side, the results obtained in our study and, in general, the complex time evolution of GRB emission
spectrum during their prompt phase, also call for the development of detailed broad-band physical models in order to 
pinpoint which processes dominate during the first instants of the GRB emisssion and to assess the contribution of
internal emission to the GeV spectrum.
For instance, our results regarding the photon spectral indices at low energies ($\alpha\sim -0.9$) and at high energies
($\gamma\sim -1.6$) are promising, since they show good agreement with prompt emission models based on fast-cooling
electron synchrotron emission with inverse Compton scatterings in Klein Nishina regime~\citep{bosnjak14}.
Dedicated simulations aiming at reproducing GRB~090926A spectral evolution in detail constitute the next step in this
direction.


\begin{acknowledgements}
The \Fermi LAT Collaboration acknowledges generous ongoing support
from a number of agencies and institutes that have supported both the
development and the operation of the LAT as well as scientific data analysis.
These include the National Aeronautics and Space Administration and the
Department of Energy in the United States, the Commissariat \`a l'Energie Atomique
and the Centre National de la Recherche Scientifique / Institut National de Physique
Nucl\'eaire et de Physique des Particules in France, the Agenzia Spaziale Italiana
and the Istituto Nazionale di Fisica Nucleare in Italy, the Ministry of Education,
Culture, Sports, Science and Technology (MEXT), High Energy Accelerator Research
Organization (KEK) and Japan Aerospace Exploration Agency (JAXA) in Japan, and
the K.~A.~Wallenberg Foundation, the Swedish Research Council and the
Swedish National Space Board in Sweden.
 
Additional support for science analysis during the operations phase is gratefully acknowledged from the Istituto Nazionale
di Astrofisica in Italy and the Centre National d'\'Etudes Spatiales in France.

\eng{The authors would like to thank J. Palmerio for his careful reading of the manuscript.}

\end{acknowledgements}

\begin{appendix}
\section{Spectral analysis results}
\label{sec:appendix}
In this section we give more information on the spectral analyses reported in~Sect.~\ref{sec:results}.

\begin{table*}[h]
  \caption{Results of the PL fits to LAT data during the \latdur time interval (from 5.5~s to 225~s post-trigger).
    \lupm{The pivot energy $E_{piv}$ in equation (\ref{eq:pl}) has been fixed to 330~MeV for Pass~7 and 240~MeV for
      Pass~8, close to the decorrelation energies.}
  }
  \label{tab:fit_latt90_pl}
  \centering
  \begin{tabular}{l c c c c}
    \hline
    \hline
    Analysis method& \multicolumn{2}{c}{Unbinned ML} & \multicolumn{2}{c}{Binned ML}\\
    \hline
    LAT data set & Pass 7   & Pass 8  & Pass 8  &  Pass 8  \\
    LAT energy range & 100~MeV-100~GeV  &  100~MeV-100~GeV &  100~MeV-100~GeV & 30~MeV-100~GeV \\
    Number of events & 319 &   464 & 464 & 1088\\
    \lupm{PL amplitude $A'$ ($\times$10$^{-4}$ cm$^{-2}$\,s$^{-1}$\,keV$^{-1}$) }& \lupm{$3.8\pm0.3$}  &\lupm{$8.7\pm0.4$}&\lupm{$8.5\pm0.4$} &\lupm{$8.1\pm0.3$} \\
    PL photon index $\gamma$& $-2.19\pm0.07$ &  $-2.25\pm0.06$ & $-2.21\pm0.06$ & $-2.20\pm0.03$\\
    $>$100~MeV flux ($10^{-5}$cm$^{-2}$\,s$^{-1}$) &  $45\pm3.1$ & $51\pm2.4$ & $50\pm2.4$ & $48\pm1.5$\\
    \hline
  \end{tabular}
\end{table*}
\begin{table*}[h] 
  \caption{Results of the CUTPL fits to LAT data during the time interval \cc (from 9.8~s to 10.5~s post-trigger).
    \lupm{The pivot energy $E_{piv}$ in equation (\ref{eq:cutpl}) has been fixed to 500~MeV, close to the decorrelation energy.}
  }
  \label{tab:fit_lat_c_cutpl}
  \centering
  \begin{tabular}{l c c c c}
    \hline
    \hline
    Analysis method& \multicolumn{2}{c}{Unbinned ML} & \multicolumn{2}{c}{Binned ML}\\
    \hline
    LAT data set  & Pass 7   & Pass 8  & Pass 8  &  Pass 8  \\
    LAT energy range & 100~MeV-100~GeV  &  100~MeV-100~GeV & 100~MeV-100~GeV & 30~MeV-100~GeV \\
    Number of events &  45  &   65 & 65 &  152\\
    \lupm{CUTPL amplitude $A'$ ($\times$10$^{-4}$ cm$^{-2}$\,s$^{-1}$\,keV$^{-1}$)} & \lupm{ $3.4\pm0.9$} & \lupm{ $3.3\pm0.7$}&  \lupm{ $1.0\pm0.6$ }&  \lupm{ $3.2\pm0.6$} \\
    CUTPL photon index $\gamma$& $-1.21\pm0.82$ &  $-1.13\pm0.68$ & $-1.11\pm0.72$ & $-1.68\pm0.22$ \\
    CUTPL \heb{folding} energy $E_f$ (GeV) &  $0.23_{-0.10}^{+0.30}$ & $0.24_{-0.09}^{+0.22}$ &  $0.26_{-0.09}^{+0.25}$ & $0.41_{-0.14}^{+0.27}$ \\
    Break significance $N_\sigma$ & 2.3  & 2.7  & 2.5  & 4.4  \\
    \hline
  \end{tabular}
\end{table*}
\begin{table*}[h] 
  \caption{Results of the Band+CUTPL fits to GBM+LAT data during the time interval \cc (from 9.8~s to 10.5~s
    post-trigger).
    \lupm{The pivot energy $E_{piv}$ in equation (\ref{eq:cutpl}) has been fixed to 1~MeV as in~\cite{090926a}.}
  }
  \label{tab:fit_joint_c_cutpl}
  \centering
  \begin{tabular}{l c c c}
    \hline
    \hline
    LAT data set & \multicolumn{1}{c}{Pass~7} & \multicolumn{2}{c}{Pass~8}\\
    \hline
    LAT energy range & 100~MeV-100~GeV   &  100~MeV-100~GeV &  30~MeV-100~ GeV \\
    Number of events &  45  &  65 & 152\\
    \lupm{Band amplitude $A_{B}$ ($\times$10$^{-2}$ cm$^{-2}$\,s$^{-1}$\,keV$^{-1}$)} &  \lupm{ $33_{-2}^{+3}$} &  \lupm{  $33_{-2}^{+4}$ } &  \lupm{  $34_{-2}^{+2}$}  \\
    Band $E_\mathrm{peak}$ (keV) & $190_{-7}^{+10}$ & $190_{-8}^{+9}$  & $189_{-9}^{+8}$  \\
    Band photon index $\alpha$ & $-0.63_{-0.15}^{+0.08}$ & $-0.62_{-0.12}^{+0.11}$ & $-0.60_{-0.14}^{+0.13}$  \\
    Band photon index $\beta$  & $-3.8_{-1.1}^{+0.4}$ & $-3.6_{-1.1}^{+0.3}$   & $-3.7_{-1.6}^{+0.5}$      \\
    \lupm{CUTPL amplitude $A'$ ($\times$10$^{-4}$ cm$^{-2}$\,s$^{-1}$\,keV$^{-1}$)} & \lupm{ $8.0_{-1.5}^{+1.1}$}  & \lupm{ $8.4_{-1.6}^{+1.1}$} & \lupm{  $8.9_{-1.4}^{+0.6}$} \\
    CUTPL photon index $\gamma$ & $-1.66_{-0.03}^{+0.05}$  &  $-1.68_{-0.03}^{+0.04}$ & $-1.68_{-0.03}^{+0.04}$ \\
    CUTPL \heb{folding} energy $E_f$ (GeV) &  $0.31_{-0.06}^{+0.08}$ &  $0.38_{-0.06}^{+0.07}$ &  $0.37_{-0.05}^{+0.06}$  \\
    Break significance $N_\sigma$ &  5.9 &  6.3 &  7.7   \\
    \hline
  \end{tabular}
\end{table*}
\begin{table*}[h] 
  \caption{Results of the Band+CUTPL fits to GBM+LAT data during the time interval \dd (from 10.5~s to 21.6~s
    post-trigger).
    \lupm{The pivot energy $E_{piv}$ in equation (\ref{eq:cutpl}) has been fixed to 1~GeV as in~\cite{090926a}.}
  } 
  \label{tab:fit_joint_d_cutpl}
  \centering
  \begin{tabular}{l c c c}
    \hline
    \hline
    LAT data set & \multicolumn{1}{c}{Pass~7} & \multicolumn{2}{c}{Pass~8}\\
    \hline
    LAT energy range & 100~MeV-100~GeV   &  100~MeV-100~GeV &  30~MeV-100~GeV \\
    Number of events &  107  &  154 & 321\\
    \lupm{Band amplitude $A_{B}$ ($\times$10$^{-2}$ cm$^{-2}$\,s$^{-1}$\,keV$^{-1}$)} & \lupm{ $9.9_{-0.6}^{+0.4}$} & \lupm{ $10.0_{-0.5}^{+0.5}$ } &  \lupm{$10.1_{-0.2}^{+0.4}$ } \\
    Band $E_\mathrm{peak}$ (keV) & $183_{-7}^{+7}$ & $182_{-6}^{+6}$  & $180_{-6}^{+5}$ \\
    Band photon index $\alpha$ & $-0.70_{-0.08}^{+0.07}$ & $-0.68_{-0.08}^{+0.07}$ & $-0.65_{-0.04}^{+0.05}$  \\
    Band photon index $\beta$  & $-2.9_{-0.2}^{+0.1}$ & $-2.9_{-0.2}^{+0.1}$   & $-2.9_{-0.3}^{+0.1}$   \\
    \lupm{CUTPL amplitude $A'$ ($\times$10$^{-10}$ cm$^{-2}$\,s$^{-1}$\,keV$^{-1}$)} &  \lupm{$4.9_{-0.6}^{+0.8}$}  & \lupm{  $6.1_{-0.8}^{+0.8}$} &  \lupm{$6.4_{-0.3}^{+0.2}$} \\
    CUTPL photon index $\gamma$ & $-1.76_{-0.03}^{+0.04}$  & $-1.77_{-0.01}^{+0.05}$ & $-1.75_{-0.03}^{+0.02}$ \\
    CUTPL \heb{folding} energy $E_f$ (GeV) &  $2.02_{-0.48}^{+0.80}$ &  $1.63_{-0.35}^{+0.53}$  &  $1.61_{-0.31}^{+0.38}$  \\
    Break significance $N_\sigma$  &  4.3 & 5.6  &  5.8   \\
    \hline
  \end{tabular}
\end{table*}
\begin{figure*}[h]
 \centering
 \vbox{
   \includegraphics[width=\linewidth]{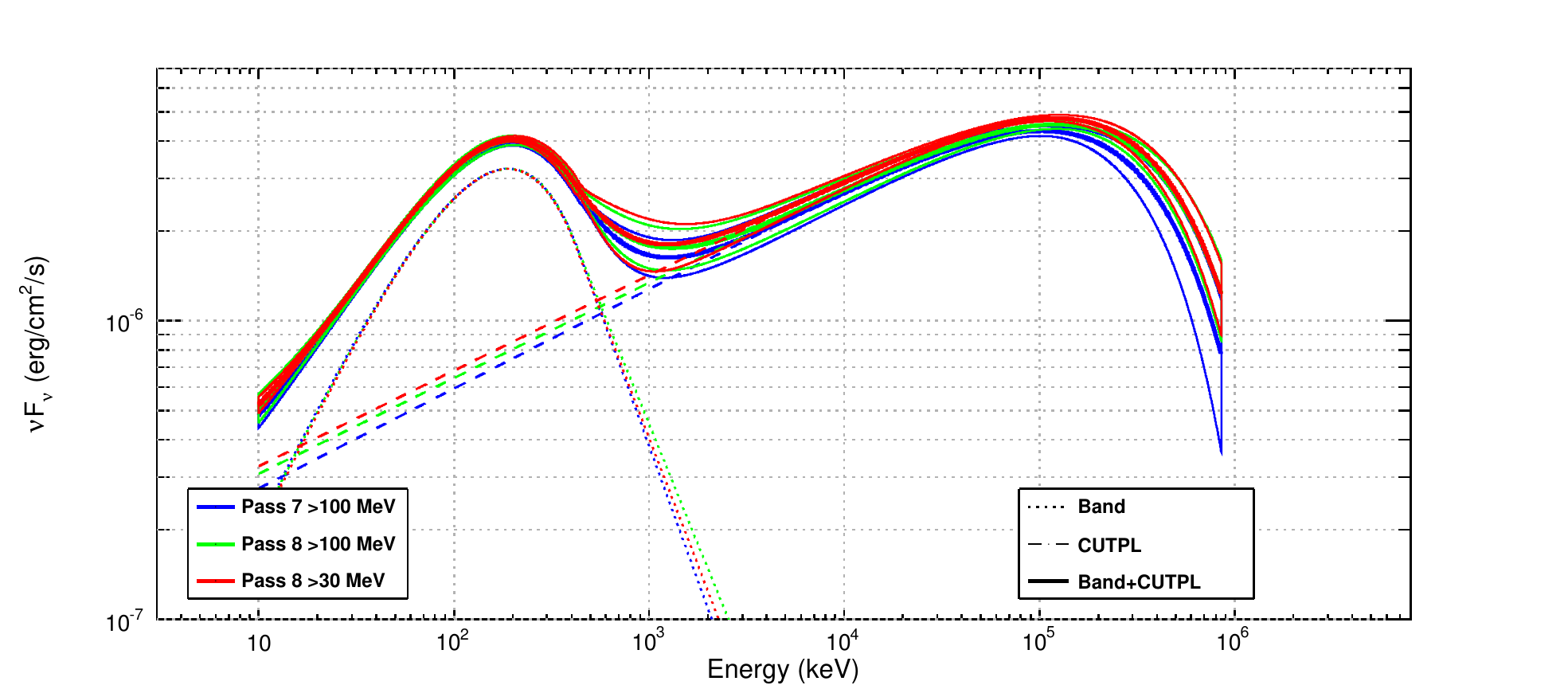}
   \includegraphics[width=\linewidth]{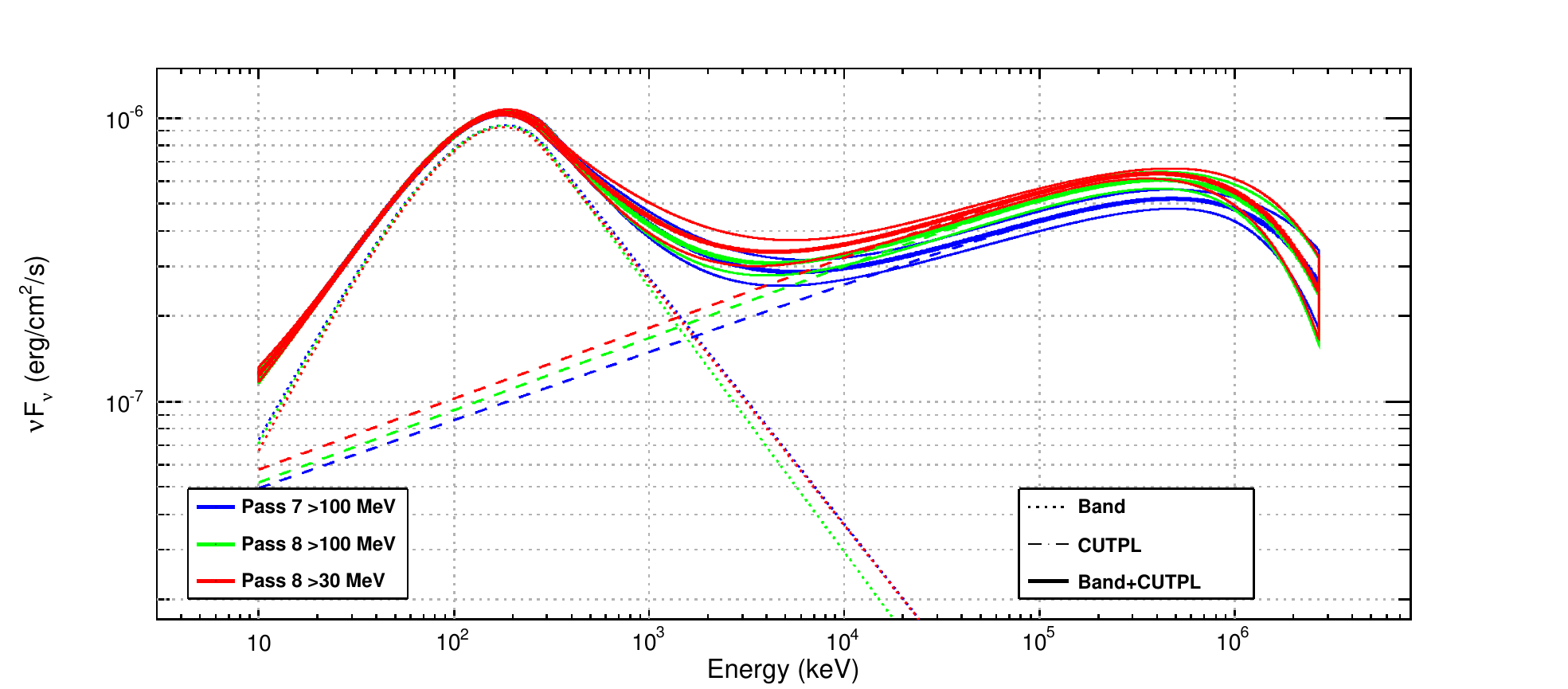}
}
 \caption{
   GRB~090926A spectral energy distributions as measured by the \Fermi GBM and LAT in time intervals \cc \corr{(top panel)} and
   \dd \corr{(bottom panel)}, using LAT Pass7 above 100~MeV, and Pass~8 data above 30~MeV and 100~MeV (see
   Tables~\ref{tab:fit_joint_c_cutpl} and \ref{tab:fit_joint_d_cutpl} for more details).
   \corr{Each solid curve represents the best fitted spectral shape (Band+CUTBPL), within a 68\% confidence level contour
     derived from the errors on the fit parameters.}
 }
 \label{fig:sed_band+cutpl}
\end{figure*}
\begin{table*}[h]
  \caption{Results of the Band+CUTBPL fits to GBM+LAT data for time sub-intervals in \cc.
  }
  \label{tab:fit_joint_c_split}
  \centering
  \begin{tabular}{l c c}
    \hline
    \hline
    Time intervals (same statistics) &  [9.80-9.98]~s    &  [9.98-10.50]~s\\
    \hline
    Number of events & 76 &   76 \\
    CUTBPL \heb{folding} energy $E_f$ (GeV)  &  $0.40_{-0.08}^{+0.10}$ &  $0.32_{-0.06}^{+0.09}$  \\
    Break significance $N_\sigma$     &  5.0              &  5.5  \\
    \hline
    \hline
    Time intervals (rise \& decay)   &  [9.80-9.94]~s    & [9.94-10.50]~s\\
    \hline
    Number of events & 49 &   103 \\
    CUTBPL \heb{folding} energy $E_f$ (GeV)  &  $0.42_{-0.10}^{+0.16}$ & $0.35_{-0.07}^{+0.08}$ \\
    Break significance $N_\sigma$     &  3.8              & 6.3 \\
    \hline
  \end{tabular}
\end{table*}
\begin{table*}[h]
  \caption{Results of the Band+CUTBPL fits to GBM+LAT data for time sub-intervals in \dd.
  }
  \label{tab:fit_joint_d_split}
  \centering
  \begin{tabular}{l c c}
    \hline
    \hline
    Time intervals (same statistics) & [10.50-12.90]~s     & [12.90-21.60]~s\\
    \hline
    Number of events & 161 &   160 \\
    CUTBPL \heb{folding} energy $E_f$ (GeV)  & $0.55_{-0.10}^{+0.13}$ & $1.43_{-0.25}^{+0.49}$  \\
    Break significance $N_\sigma$     & 4.3                  & 5.1  \\
    \hline
    \hline
    Time intervals (rise \& decay)   & [10.50-11.73]~s     & [11.73-21.60]~s\\
    \hline
    Number of events & 81 &   240 \\
    CUTBPL \heb{folding} energy $E_f$ (GeV)  & $0.45_{-0.10}^{+0.17}$ & $1.85_{-0.30}^{+0.75}$ \\
    Break significance $N_\sigma$     & 5.1                  & 4.4 \\
    \hline
  \end{tabular}
\end{table*}
\end{appendix}


%
%

\bibliographystyle{aa} 
\bibliography{biblio.bib} 
\end{document}